\documentclass[aps,pra,superscriptaddress,twocolumn]{revtex4}
\usepackage[dvipdfmx]{graphicx}
\usepackage{graphicx}
\usepackage{color}
\usepackage{textcomp}
\usepackage{amsmath,amssymb,amsthm}
\usepackage{bm}
\usepackage{comment}
\usepackage{ulem}
\usepackage{grffile}

\begin{document}


\title{Nonequilibrium steady states of Bose-Einstein condensates with a local particle loss in double potential barriers 
}

\author{Masaya Kunimi}
\email{E-mail: kunimi@ims.ac.jp}
\thanks{Present address : Department of Photo-Molecular Science, Institute for Molecular Science, National Institutes of Natural Sciences, Myodaiji, Okazaki 444-8585, Japan}
\affiliation{Yukawa Institute for Theoretical Physics, Kyoto University, Kyoto 606-8502, Japan}
\author{Ippei Danshita}
\affiliation{Department of Physics, Kindai University, Higashi-Osaka, Osaka 577-8502, Japan}




\date{\today}

\begin{abstract}
We investigate stability of non-equilibrium steady states of Bose-Einstein condensates with a local one-body loss in the presence of double potential barriers. We construct an exactly solvable mean-field model, in which the local loss and the potential barriers take the form of a delta function. Using the exact solutions of our model, we show that there are parameter regions in which two steady-state solutions are dynamically stable, i.e., the model exhibits bistability. We also find that unidirectional hysteresis phenomena appear when the local-loss rate is varied in some parameter region.
\end{abstract}

\maketitle

\section{Introduction}\label{sec:Introduction}

Ultracold gases are well known as coherent quantum systems with high controllability \cite{Bloch2008}. Ultracold gases are confined in a vacuum chamber by using magnetic fields or laser beams such that they are well decoupled from environments. This means that ultracold gases are regarded as isolated quantum systems \cite{Polkovnikov2011}. Many interesting phenomena have been studied in the context of isolated quantum systems, such as thermalization \cite{Rigol2008,Polkovnikov2011,Gring2012,Kaufman2016,Mori2018} and many-body localization \cite{Schreiber2015,Smith2016,Choi2016,Nandkishore2015,Altman2015,Abanin2018a}.

Recent technological advances in ultra-cold atom experiments allow us to introduce couplings to the environment, namely, dissipation, in a well-controlled manner \cite{Syassen2008,Barontini2013,Yan2013,Labouvie2015,Patil2015,Labouvie2016,Luschen2017,Tomita2017}. This means that we can switch ultracold gases from isolated systems to controllable open many-body quantum systems \cite{Diehl2008,Verstraete2009,Daley2014,Ashida2016}. The dissipation can be regarded as continuous measurements. When the dissipation is strong compared to other energy scales of the systems, quantum Zeno effects occur \cite{Misra1977}, which suppress coherent processes such as tunneling. These effects have been observed in ultracold-gas experiments \cite{Syassen2008,Mark2012,Yan2013,Barontini2013,Labouvie2015,Patil2015,Labouvie2016,Tomita2017}. It is also noteworthy that the controllable dissipations provide us new possibilities for exploring novel quantum systems, such as $\mathcal{PT}$ symmetric systems \cite{Bender1998,Ruter2010,Konotop2016,Xiao2017,Li2019} and non-Hermitian quantum systems \cite{Bender2007,El-Ganainy2018,Gong2018}. 

Recently, the experimental group at Technische Universitat Kaiserslauten observed bistability in a Bose-Einstein condensate (BEC) with a local particle loss confined in a one-dimensional optical lattice~\cite{Labouvie2016}. The local particle loss can be realized by focusing an electron beam on the central site of the optical lattice. They prepared two different initial conditions. One is that the central site of the optical lattice is occupied by the particles and the other is that the central site is almost empty. Measuring the particle number of the central site by using scanning electron microscopy techniques, they observed two different stable states. In the small (strong) dissipation regime, the occupied (empty) state is realized regardless of the initial conditions. On the other hand, at the intermediate dissipation strength, the two different stable states are realized depending on the initial states. This means that the system exhibits bistability.

This experiment can be understood as a problem of stability of supercurrents under particle losses. Because the local particle loss induces a density difference between the central site and the others, the supercurrent flows from the surrounding sites into the central sites. The results observed in the experiment indicate that particle losses produce nontrivial effects on superfluidity. In fact, our previous work also showed that global three-body losses induce supercurrent decay in a ring trap \cite{Kunimi2019}. 

In previous theoretical studies~\cite{Brazhnyi2009,Sels2018a}, it has been shown that in the absence of optical lattice potentials, which are described by a real-number external field in the Gross-Pitaevskii (GP) equation, the system does not exhibit a discontinuous jump in the density under a local one-body loss associated with the bistability when the strength of the dissipation is varied. This is contrary to an experiment~\cite{Labouvie2016}, in which an optical lattice potential is present. In this work,  we construct a simple model that is analytically solvable and exhibits the discontinuous jump associated with bistability. Specifically, we use a one-dimensional GP equation with a local one-body loss and double potential barriers, which are,  respectively, described by pure imaginary and real delta function potentials. On the basis of semi-analytical solutions of our model, we indeed show that the inclusion of the double potential barriers leads to the emergence of bistability accompanied by the discontinuous jump. In addition, we find unidirectional hysteresis phenomena in our systems. These phenomena are called anomalous hysteresis~\cite{Yamamoto2012,Yamamoto2013_2,Yamamoto2013}.

This paper is organized as follows. In Sec.~\ref{sec:model}, we explain the problem that we consider and its formulation based on a dissipative GP equation, which describes a BEC with a local particle loss. In Sec.~\ref{subsec:absence_of_pinning_potential}, using the exact solution of the GP equation, we briefly review important properties of the BEC in the absence of double potential barriers. In Sec.~\ref{subsec:presence_of_pinning_potential}, we obtain exact solutions of the GP equation in the presence of double potential barriers in order to discuss the stability of nonequilibrium steady states of the BEC. In Sec.~\ref{subsec:Anomalous hysteresis}, we show that our system exhibits anomalous hysteresis phenomena. In Sec.~\ref{sec:summary}, we summarize our results. In the Appendixes, we explain how to perform the stability analysis of stationary solutions of the GP equation and the details of the derivations of the exact solution of the GP equation.

\section{Model}\label{sec:model}
\begin{figure}[t]
\centering
\includegraphics[width=8.6cm]{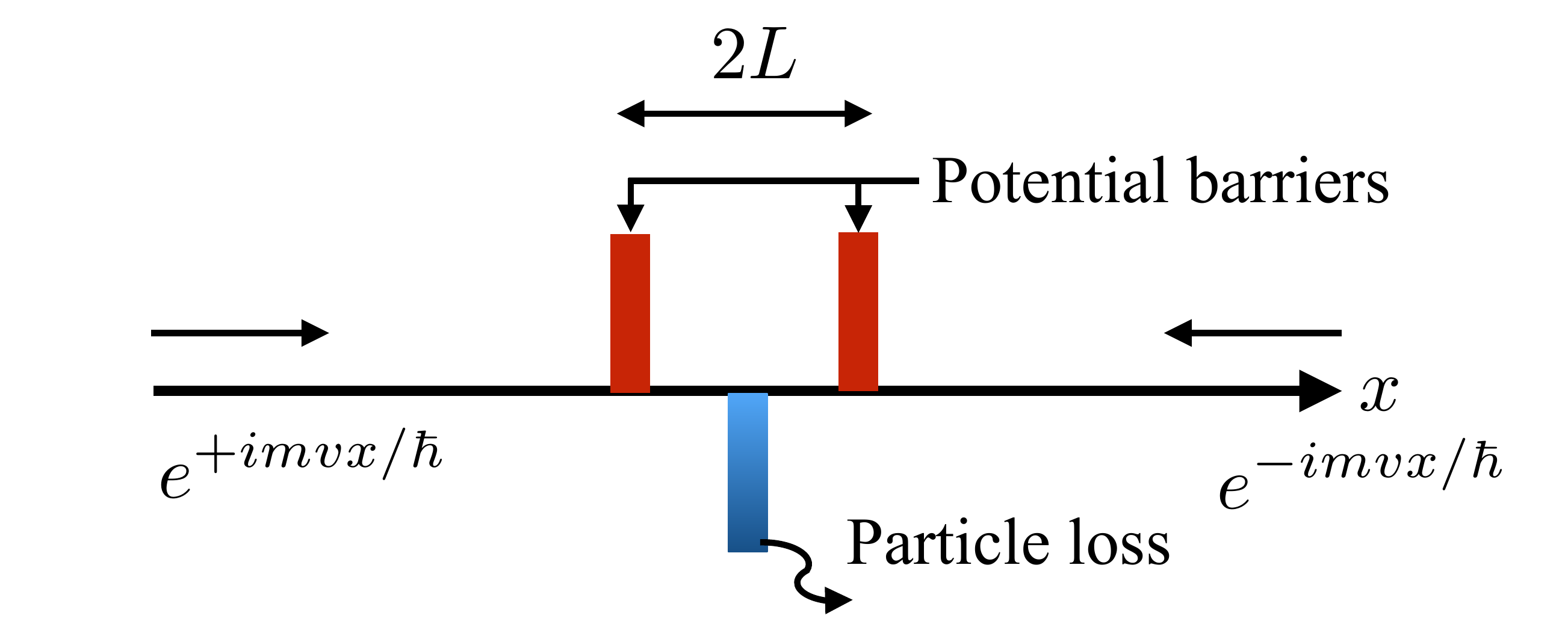}
\caption{Schematic of our setup.}
\label{fig:schematic_picture_setup}
\end{figure}%

In this paper, we consider a one-dimensional GP equation with a local one-body loss term and double-potential-barrier terms,
\begin{align}
i\hbar\frac{\partial}{\partial t}\psi(x, t)&=\left[-\frac{\hbar^2}{2M}\frac{\partial^2}{\partial x^2}+U(x)+g|\psi(x, t)|^2\right]\psi(x, t),\label{eq:time-dependent_GP_eq}\\
U(x)&\equiv -\frac{i\hbar\gamma_0}{2}\delta(x)+U_0[\delta(x-L)+\delta(x+L)],\label{eq:definition_of_potential}
\end{align}
where $M$ is the mass of the atom, $g>0$ is the two-body interaction coefficient, and $\psi(x, t)$ is the order parameter of the BEC. 
The dissipation term takes the form of the delta function localized at $x=0$ and $\gamma_0\ge0$ is the strength of the dissipation. 
The two potential barriers located at $x = \pm L$ are added to mimic the density dips near the local loss created by the optical lattice in the experiment \cite{Labouvie2016}. Their strength is denoted $U_0\ge0$. This dissipative GP equation can be derived by the mean-field approximation of the Lindblad equation with the local one-body loss term (see details in the Supplemental Material of Ref.~\cite{Barontini2013}).

In Sec.~\ref{sec:results}, we consider the stability of nonequilibrium steady states of a BEC, in which a stationary supercurrent flows into the location of the particle loss. Such states are represented as solutions of the time-independent GP equation, which is derived by inserting $\psi(x, t)=\Psi(x)e^{-i\mu t/\hbar}$ into Eq.~(\ref{eq:time-dependent_GP_eq}),
\begin{align}
\left[-\frac{\hbar^2}{2M}\frac{d^2}{d x^2}+U(x)-\mu+g|\Psi(x)|^2\right]\Psi(x)=0,\label{eq:time-independent_GP}
\end{align}
where $\mu$ is the chemical potential.

We set the boundary condition at $x\to\pm\infty$ as (see also Fig.~\ref{fig:schematic_picture_setup})
\begin{align}
\Psi(x)\xrightarrow{x\to\pm\infty}\sqrt{n_{\infty}}e^{-i M v_{\infty}|x|/\hbar}e^{i\varphi_{\pm}},\label{eq:boundary_condition_at_infinity}
\end{align}
where $n_{\infty}\ge 0$ is the mean particle density at $|x|\rightarrow\infty$, $v_{\infty}\ge0$ is the magnitude of the flow velocity at $|x|\rightarrow\infty$, and $\varphi_{\pm}$ is the phase. From this boundary condition, we obtain the chemical potential:
\begin{align}
\mu&=g n_{\infty}+\frac{1}{2}Mv_{\infty}^2.\label{eq:chemical_potential_for_all_state}
\end{align}
The velocity $v_{\infty}$ is determined by the boundary conditions due to the delta functions, which are given by
\begin{align}
&\Psi(\pm L+0)=\Psi(\pm L-0),\quad \Psi(+0)=\Psi(-0),\label{eq:boundary_condition_for_continuity}\\
&\frac{\hbar^2}{2M}\left[\left.\frac{d\Psi(x)}{dx}\right|_{x=\pm L+0}-\left.\frac{d\Psi(x)}{d x}\right|_{x=\pm L-0}\right]=U_0\Psi(\pm L),\label{eq:boundary_condition_pinning_potential}\\
&\frac{\hbar^2}{2M}\left[\left.\frac{d\Psi(x)}{dx}\right|_{x=+0}-\left.\frac{d\Psi(x)}{d x}\right|_{x=-0}\right]=-\frac{i\hbar\gamma_0}{2}\Psi(0).\label{eq:boundary_condition_loss_term}
\end{align}

We check the stability of the obtained stationary solutions by the numerical simulations of the time-dependent GP equation. For the details see Appendix \ref{app:stability_analysis}.

At the end of this section, we remark on a crucial difference between our model and the actual experimental setup. In our setup, the particles are lost at the origin and provided at $|x|\rightarrow \infty$ [see Eq.~(\ref{eq:boundary_condition_at_infinity})]. This fact can be easily seen by writing down the equation of continuity,
\begin{align}
\frac{\partial}{\partial t}n(x, t)&=-\frac{\partial}{\partial x}J(x, t)-\gamma_0\delta(x)n(x, t),\label{eq:equation_of_continuity}\\
n(x, t)&\equiv |\psi(x, t)|^2,\label{eq:definition_of_density}\\
J(x, t)&\equiv -\frac{i\hbar}{2M}\left[\psi^{\ast}(x, t)\frac{\partial}{\partial x}\psi(x, t)-{\rm c.c.}\right],\label{eq:definition_of_current_density}
\end{align}
where $n(x, t)$ and $J(x, t)$ are the particle density and the current density, respectively. Integrating Eq.~(\ref{eq:equation_of_continuity}) over $(-\infty,+\infty)$ yields 
\begin{align}
\frac{d}{d t}N(t)&=-[J(+\infty,t)-J(-\infty,t)]-\gamma_0n(0,t),\label{eq:change_of_total_particle}
\end{align}
where $N(t)\equiv \int^{+\infty}_{-\infty} dx [n(x, t)-n_{\infty}]$ is the total particle number difference at time $t$ \cite{Note1}. The first and second terms on the right-hand side of Eq.~(\ref{eq:change_of_total_particle}) represent the gain of the particles from the boundaries and the third one represents the loss of the particles at $x=0$. This equation shows that nonequilibrium steady states can be realized when the loss and gain of the particles are balanced.

In the experiment, the BEC is confined in the trap potential with the local particle loss. Because there is no particle source, in contrast to our theoretical setup, the total particle number in the trap monotonically decreases. Hence, strictly speaking, the stationary states cannot exist except in a vacuum state (no particle in the trap). However, according to the inset in Fig.~2 (a) in Ref.~\cite{Labouvie2016}, we can see that the particle number at the central site is almost stationary over the time scale $40$-$60\;{\rm ms}$. In this time scale, the particle loss and the hopping from the adjacent sites to the central site are balanced. As long as we focus on the vicinity of the central site, the systems can be approximated as nonequilibrium steady states. Stationary states in our model correspond to these nonequilibrium steady states.

Another difference is the width of the local dissipation term. As described above, we assume that the local dissipation is given by the delta function. This treatment can be justified when the width of the dissipation is much smaller than the healing length. However, in the experiment, the width of the dissipation is about $O(0.1\mu{\rm m})$ \cite{Barontini2013}. Because the healing length of the experiment is $O(0.1\mu{\rm m})$, the dissipation in the experiment cannot be regarded as the delta function. We also remark on the effects of the finite width in Sec.~\ref{subsec:presence_of_pinning_potential}.

\section{Results}\label{sec:results}

\subsection{In the absence of double potential barriers}\label{subsec:absence_of_pinning_potential}
For the reader's convenience we first review exact solutions in the absence of double potential barriers, which have been derived in some previous works \cite{Brazhnyi2009,Sels2018a} before showing our results.

There are three kinds of exact solutions in the absence of potential barriers ($U_0=0$). One is a plane-wave (PW) solution:
\begin{align}
\Psi_{\rm PW}(x)&=\sqrt{n_{\infty}}e^{-i M v_{\infty}|x|/\hbar},\label{eq:plane_wave_solution_wave_function}\\
v_{\infty}&=\frac{\gamma_0}{2}.\label{eq:velocity_plane_wave}
\end{align} 
The second is a dark soliton (DS) solution,
\begin{align}
\Psi_{\rm DS}(x)&=\sqrt{n_{\infty}}\tanh(x/\xi),\label{eq:dark_soliton_wave_function}\\
v_{\infty}&=0,\label{eq:velocity_dark_soliton}
\end{align}
where $\xi\equiv \hbar/\sqrt{M g n_{\infty}}$ is the healing length. The last one is a gray soliton (GS) solution:
\begin{align}
\Psi_{\rm GS}(x)&=\sqrt{n_{\infty}}e^{-i M v_{\infty}|x|/\hbar}\left[i\frac{v_{\infty}}{v_{\rm s}}+f(x)\right],\label{eq:wave_function_gray_soliton}\\
f(x)&\equiv \sqrt{1-\left(\frac{v_{\infty}}{v_{\rm s}}\right)^2}\tanh\left[\sqrt{1-\left(\frac{v_{\infty}}{v_{\rm s}}\right)^2}\frac{|x|}{\xi}\right],\label{eq:definition_of_function_f}\\
v_{\infty}&=\frac{2v_{\rm s}^2}{\gamma_0},\label{eq:velocity_gray_soliton}
\end{align}
where $v_{\rm s}\equiv \sqrt{g n_{\infty}/M}$ is the sound velocity. We can easily check that these expressions satisfy the GP equation (\ref{eq:time-independent_GP}). We note that the PW and DS solutions exist for arbitrary parameters and the GS solution exists for $\gamma_0> 2v_{\rm s}$.

\begin{figure}[t]
\centering
\includegraphics[width=8cm]{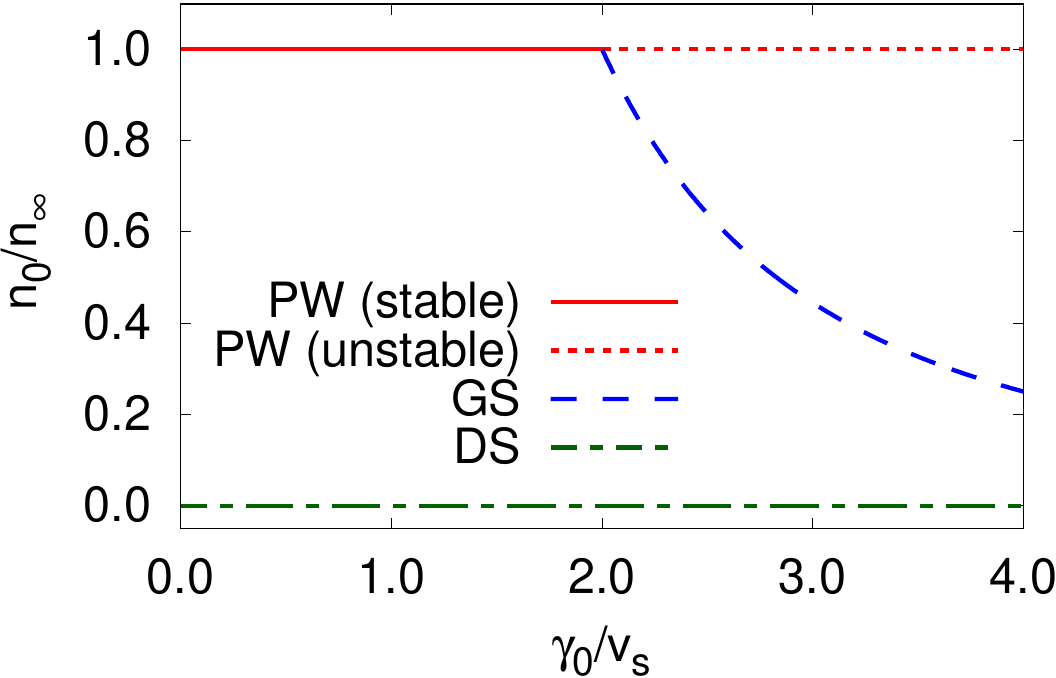}
\caption{Density at the origin as a function of the dissipation strength. The solid red, dotted red, dashed blue, and dashed-dotted green lines represent the stable PW solutions, unstable PW solutions, GS solutions, and DS solutions, respectively.}
\label{fig:density_at_origin_U0}
\end{figure}%
\begin{figure}[t]
\centering
\includegraphics[width=8cm]{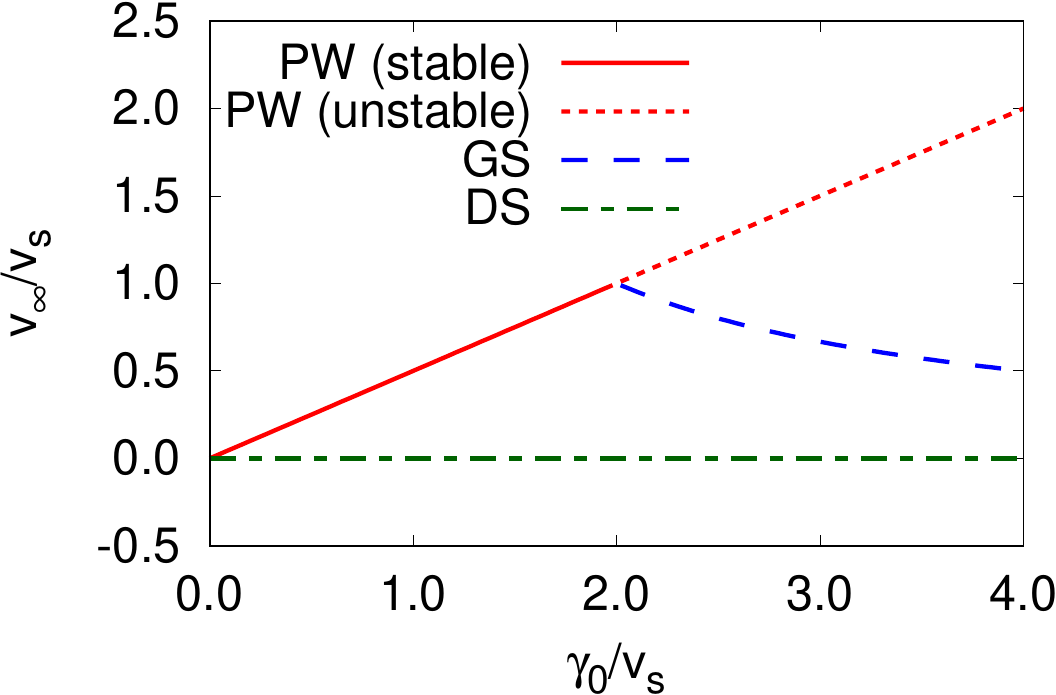}
\caption{Magnitude of the flow velocity at infinity as a function of the dissipation strength. The solid red, dotted red, dashed blue, and dashed-dotted green lines represent the stable PW solutions, unstable PW solutions, GS solutions, and DS solutions, respectively.}
\label{fig:velocity_vs_gamma_U0}
\end{figure}%

Here, we define the density at the origin as $n_0\equiv |\Psi(x=0)|^2$, which corresponds to the density at the central site in the experiment \cite{Labouvie2016}. We plot $n_0$ as a function of $\gamma_0$ in Fig.~\ref{fig:density_at_origin_U0}, which we call an $n_0$-$\gamma_0$ diagram. This result shows that the system exhibits bistability in the whole $\gamma_0$ region. For $\gamma_0\le 2v_{\rm s}$ the PW and the DS states are stable, and for $\gamma_0>2v_{\rm s}$ the GS and the DS states are stable. We can see the unstable PW states for $\gamma_0>2v_{\rm s}$. This can be understood by the velocity shown in Fig.~\ref{fig:velocity_vs_gamma_U0}. The velocity of the PW state is given by Eq.~(\ref{eq:velocity_plane_wave}), which is proportional to the dissipation strength $\gamma_0$. When the velocity exceeds the sound velocity, which is the Landau critical velocity of uniform superfluids \cite{Landau1941,Wu2003}, the PW state becomes energetically unstable. 

The GS state emerges at $\gamma_0=2v_{\rm s} \;(v_{\infty}=v_{\rm s})$. The velocity of the GS state is a monotonically decreasing function of $\gamma_0$ [see Eq.~(\ref{eq:velocity_gray_soliton})]. We can interpret this behavior as follows. Suppose that we start with the PW state at $\gamma_0=0$.  When we increase the dissipation strength from $\gamma_0=0$, the superflow velocity becomes high and then reaches the Landau critical velocity. Finally, the PW states become unstable and bifurcate into the unstable PW branch and the stable GS branch.

In the DS states, the density at the origin is always $0$. This means that the DS states do not feel the dissipation. In fact, boundary condition (\ref{eq:boundary_condition_loss_term}) is satisfied in the DS solution (\ref{eq:dark_soliton_wave_function}), for the whole $\gamma_0$ region. Therefore, the DS states always exist regardless of the dissipation strength.

\subsection{In the presence of double potential barriers}\label{subsec:presence_of_pinning_potential}
Here, we show the results in the presence of double potential barriers. We assume that a functional form of the stationary solution is given by an even function or an odd function. Owing to this assumption, it is sufficient to consider only the $x\ge 0$ region. Because the potentials are only the delta function type, we can separately solve the GP equation in an inside region ($0\le x\le L$) and an outside region ($x>L$). After obtaining the solutions of each region, we connect them by using the boundary conditions (\ref{eq:boundary_condition_for_continuity}), (\ref{eq:boundary_condition_pinning_potential}), and (\ref{eq:boundary_condition_loss_term}). Such techniques for solving the GP equation with delta-function potentials have been developed in the context of Josephson junction systems \cite{Baratoff1970,Sols1994,Hakim1997,Kovrizhin2001,Pham2002,Pavloff2002,Kagan2003,Astrakharchik2004,Seaman2005,Bilas2005,Danshita2006,Danshita2007,Watanabe2009,Sykes2009,Takahashi2009,Piazza2010,Kato2010,Watabe2013,Cominotti2014}. For convenience, we introduce the following variables
\begin{align}
\Psi(x)&\equiv
\begin{cases}
\vspace{0.4em}\Psi_{\rm in}(x)\equiv \sqrt{n_{\rm in}(x)}e^{i\varphi_{\rm in}(x)},\quad \text{for }0\le x\le L, \\
\Psi_{\rm out}(x)\equiv \sqrt{n_{\rm out}(x)}e^{i\varphi_{\rm out}(x)},\quad\text{for } x>L,
\end{cases}
\label{eq:definition_of_inside_and_outside_solution}
\end{align}

First, we consider the even-function case. The solution of the outside region is given by
\begin{align}
\frac{n_{\rm out}(x)}{n_{\infty}}&=\left(\frac{v_{\infty}}{v_{\rm s}}\right)^2+\left[1-\left(\frac{v_{\infty}}{v_{\rm s}}\right)^2\right]\notag \\
&\times \tanh^2\left[\sqrt{1-\left(\frac{v_{\infty}}{v_{\rm s}}\right)^2}\frac{x-L+ x_{+}}{\xi}\right],\label{eq:solution_outside}\\
\varphi_{\rm out}(x)&=\varphi_L-\frac{M v_{\infty} (x-L)}{\hbar}\notag \\
&-\tan^{-1}\left[\frac{G(x+ x_+)}{v_{\infty}/v_{\rm s}}\right]+\tan^{-1}\left[\frac{G(L+ x_+)}{v_{\infty}/v_{\rm s}}\right],\label{eq:simple_form_of_phase}\\
\frac{x_+}{\xi}&\equiv \frac{1}{\sqrt{1-\left(\dfrac{v_{\infty}}{v_{\rm s}}\right)^2}}\notag \\
&\times \tanh^{-1}\left[\sqrt{\frac{n_L/n_{\infty}-(v_{\infty}/v_{\rm s})^2}{1-(v_{\infty}/v_{\rm s})^2}}\right],\label{eq:definition_of_x_plus_outside}\\
G(x)&\equiv\sqrt{1-\left(\frac{v_{\infty}}{v_{\rm s}}\right)^2}\tanh{\left[\sqrt{1-\left(\frac{v_{\infty}}{v_{\rm s}}\right)^2}\frac{x-L}{\xi}\right]},\label{eq:definition_of_function_G}
\end{align}
where $\varphi_L\equiv \varphi(x=L)$ and $n_L\equiv n(x=L)$ are determined using the boundary conditions below. $v_{\infty}$ is given by
\begin{align}
v_{\infty}&=\frac{1}{2}\frac{n_0}{n_{\infty}}\gamma_0.\label{eq:expression_v_infty}
\end{align}
This relation can be derived by using the assumption of an even function, the expression of the current density, and the boundary conditions (\ref{eq:boundary_condition_loss_term}). The details of the derivation of the outside solution and Eq.~(\ref{eq:expression_v_infty}) are summarized in Appendix \ref{app:derivation}.

In the inside region, we find four types of inside solutions. However, only two solutions appear in the parameter regions of our interest, where $0\le \gamma_0/v_{\rm s}\le 4$ and $0\le n_0/n_{\infty}\le 1$. Then we consider two types of solutions:
\begin{align}
\frac{n_{\rm in}^{(1)}(x)}{n_{\infty}}&=A-\left(A-\frac{n_0}{n_{\infty}}\right){\rm nd}^2(\Delta^{1/4}x/\xi| m_1),\label{eq:density_solution1}\\
\varphi_{\rm in}^{(1)}(x)&=-\frac{1}{2A}\frac{n_0}{n_{\infty}}\frac{\gamma_0}{v_{\rm s}}\frac{x}{\xi}-\frac{1}{2\Delta^{1/4}}\frac{\gamma_0}{v_{\rm s}}\frac{A-n_0/n_{\infty}}{A}\notag \\
&\times \Pi[m_1A/(n_0/n_{\infty}); {\rm am}(\Delta^{1/4}x/\xi|m_1) | m_1],\label{eq:phase_for_inside_solution_1}\\
m_1&\equiv 1-\frac{A-n_0/n_{\infty}}{\sqrt{\Delta}},\label{eq:m1_solution1}\\
\frac{n_{\rm in}^{(2)}(x)}{n_{\infty}}&=\frac{n_0}{n_{\infty}}+\left(B-\frac{n_0}{n_{\infty}}\right){\rm sn}^2(\Delta^{1/4}x/\xi | m_2),\label{eq:density_solution2}\\
\varphi_{\rm in}^{(2)}(x)&=-\frac{1}{2\Delta^{1/4}}\frac{\gamma_0}{v_{\rm s}}\notag \\
&\times \Pi\left[\left.\frac{B-n_0/n_{\infty}}{n_0/n_{\infty}}; {\rm am}(\Delta^{1/4}x/\xi | m_2)\right|m_2\right],\label{eq:phase_for_solution_2_inside}\\
m_2&\equiv \frac{B-n_0/n_{\infty}}{A-n_0/n_{\infty}},\label{eq:m2_solution2}
\end{align}
where we have set the origin of the phase as $\varphi_{\rm in}^{(i)}(x=0)=0$ and used the Jacobi elliptic functions ${\rm sn}(x|m)$ and ${\rm nd}(x|m)\equiv 1/{\rm dn}(x|m)$, the incomplete elliptic integral of the third kind $\Pi(n; \phi|m)$, and the Jacobi amplitude function ${\rm am}(x|m)$. The notations for the Jacobi elliptic functions and the elliptic integrals follow by Abramowitz and Stegun \cite{Abramowitz_Stegun}. We also used the following quantities:
\begin{align}
A&\equiv \frac{1}{2}\left[2+\frac{1}{4}\left(\frac{\gamma_{0}}{v_{\rm s}}\right)^2\left(\frac{n_0}{n_{\infty}}\right)^2-\frac{n_0}{n_{\infty}}+\sqrt{\Delta}\right],\label{eq:expression_of_A_fixed}\\
B&\equiv \frac{1}{2}\left[2+\frac{1}{4}\left(\frac{\gamma_{0}}{v_{\rm s}}\right)^2\left(\frac{n_0}{n_{\infty}}\right)^2-\frac{n_0}{n_{\infty}}-\sqrt{\Delta}\right],\label{eq:expression_of_B_fixed}\\
\Delta&\equiv \left[\frac{n_0}{n_{\infty}}-2-\frac{1}{4}\left(\frac{\gamma_{0}}{v_{\rm s}}\right)^2\left(\frac{n_0}{n_{\infty}}\right)^2\right]^2-\left(\frac{\gamma_{0}}{v_{\rm s}}\right)^2\frac{n_0}{n_{\infty}}.\label{eq:definition_of_Delta}
\end{align}
From the above results and boundary condition (\ref{eq:boundary_condition_for_continuity}), $n_L$ and $\varphi_L$ are determined by
\begin{align}
n_L&=n_{\rm in}^{(i)}(x=L), \quad \varphi_L=\varphi_{\rm in}^{(i)}(x=L).\label{eq:definition_of_nL_and_phiL}
\end{align}

Next, we consider the odd-function case. From this assumption, we obtain $\Psi(x=0)=0$. This means that the odd-function solution does not depend on $\gamma_0$ (see the descriptions of the DS in Sec.~\ref{subsec:absence_of_pinning_potential}). From the equation of continuity, the current density is independent of $x$. In this case, $J(x)=0$ because $\Psi(0)=0$. Therefore, the odd-function solution does not carry a supercurrent and we can take $\Psi(x)$ as a real function without loss of generality. The solution is given by
\begin{align}
\Psi_{\rm out}(x)&=\sqrt{n_{\infty}}\tanh\left(\frac{x-L+x_0}{\xi}\right)e^{i\varphi_0},\label{eq:tanh_solution_for_odd_function}\\
\Psi_{\rm in}(x)&=\sqrt{n_{\infty}}\sqrt{\frac{2m_0}{1+m_0}}{\rm sn}\left(\left.\sqrt{\frac{2}{1+m_0}}\frac{x}{\xi}\right|m_0\right),\label{eq:inside_solution_for_odd_function}
\end{align}
where $\varphi_0=0\text{ or }\pi$, and $x_0$ and $m_0$ are constants. $\varphi_0$ and $x_0$ are determined by boundary condition (\ref{eq:boundary_condition_for_continuity}):
\begin{align}
\tanh\left(\frac{x_0}{\xi}\right)e^{i\varphi_0}&=\sqrt{\frac{2m_0}{1+m_0}}{\rm sn}\left(\left.\sqrt{\frac{2}{1+m_0}}\frac{L}{\xi}\right|m_0\right).\label{eq:matching_condition_equation_odd_function}
\end{align}

Although the functional forms of the exact solution have been derived, $n_0$ (for the even-function case) and $m_0$ (for the odd-function case) have not been determined yet. These variables can be determined by solving boundary condition (\ref{eq:boundary_condition_pinning_potential}). Unfortunately, we cannot solve Eq.~(\ref{eq:boundary_condition_pinning_potential}) analytically. We solve Eq.~(\ref{eq:boundary_condition_pinning_potential}) numerically. The details of the derivations of these solutions are reported in Appendixes \ref{app:derivation} and \ref{app:derivation_odd}.

Here, we remark on the range of $L$. From Eqs.~(\ref{eq:density_solution1}) and (\ref{eq:density_solution2}), we can find that the inside solutions have periodicity $2K(m_1)\xi/\Delta^{1/4}$ and $2K(m_2)\xi/\Delta^{1/4}$ due to the properties of the Jacobi elliptic functions, where $K(\cdot)$ is the complete elliptic integral of the first kind. If $L$ is much larger than these periods, we can expect that there are solutions that oscillate multiple times in the inside region. To avoid the complexity of the problem, we restrict the range of $L$ to $0\le L\lesssim 3.3$, which means that the number of oscillations in the inside region is less than 1.

\begin{figure}[t]
\centering
\includegraphics[width=8.5cm]{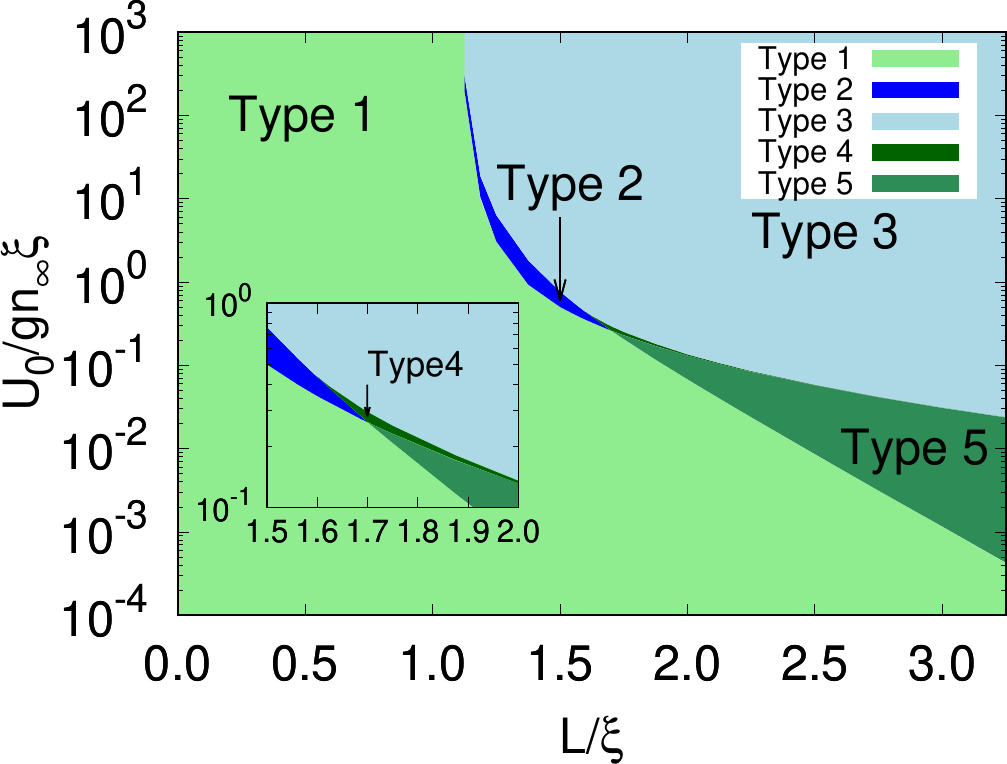}
\caption{State phase diagram of $U_0$ vs $L$. Inset:  Magnified view around the type 4 region.}
\label{fig:state_phase_diagram}
\end{figure}%

In the presence of double potential barriers, we find five types of $n_0$-$\gamma_0$ diagrams. The parameter region for the $n_0$-$\gamma_0$ diagrams is shown in Fig.~\ref{fig:state_phase_diagram}. 

\begin{figure}[t]
\centering
\includegraphics[width=8.5cm]{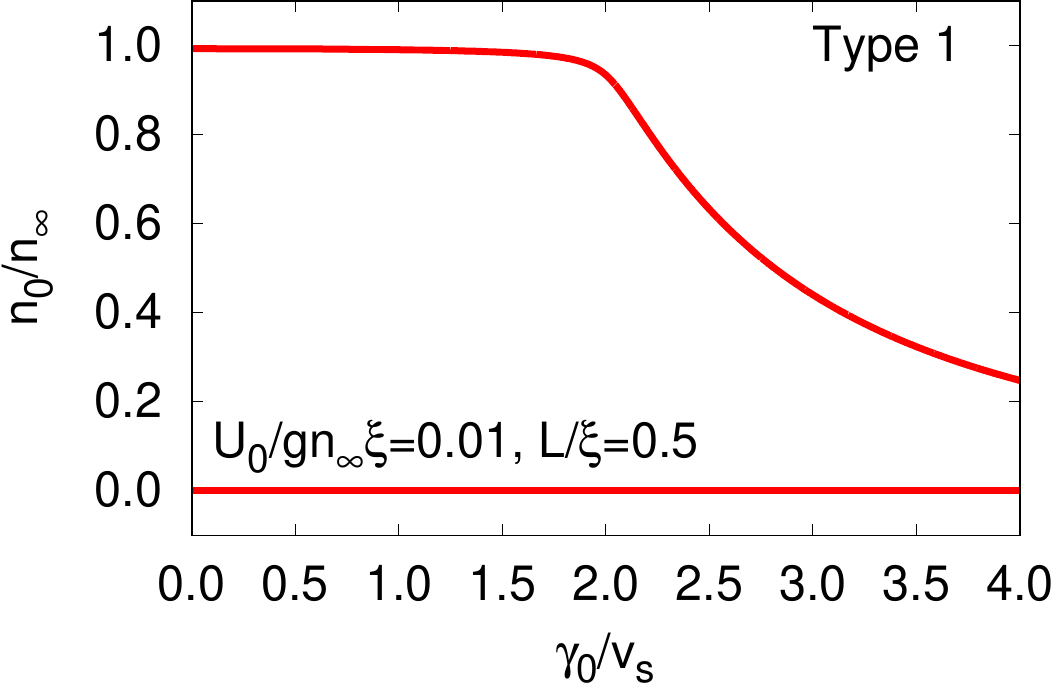}
\caption{$n_0$-$\gamma_0$ diagram for $U_0=0.01g n_{\infty}\xi$ and $L=0.5\xi$.}
\label{fig:n0-gamma0_U0.01_L0.5}
\end{figure}

A typical type 1diagram is shown in Fig.~\ref{fig:n0-gamma0_U0.01_L0.5}. In type 1, we have two stable branches. One is the even function (upper branch) and the other is the odd function (lower branch). The type 1 solution tends to exist in a region where $U_0$ is small. This means that type 1 can be interpreted as perturbed $U_0=0$ states. In fact, the $n_0$-$\gamma_0$ diagram in Fig.~\ref{fig:state_phase_diagram} is similar to that in Fig~\ref{fig:density_at_origin_U0} except for the existence of the unstable PW branch. 

\begin{figure}[t]
\centering
\includegraphics[width=8.5cm]{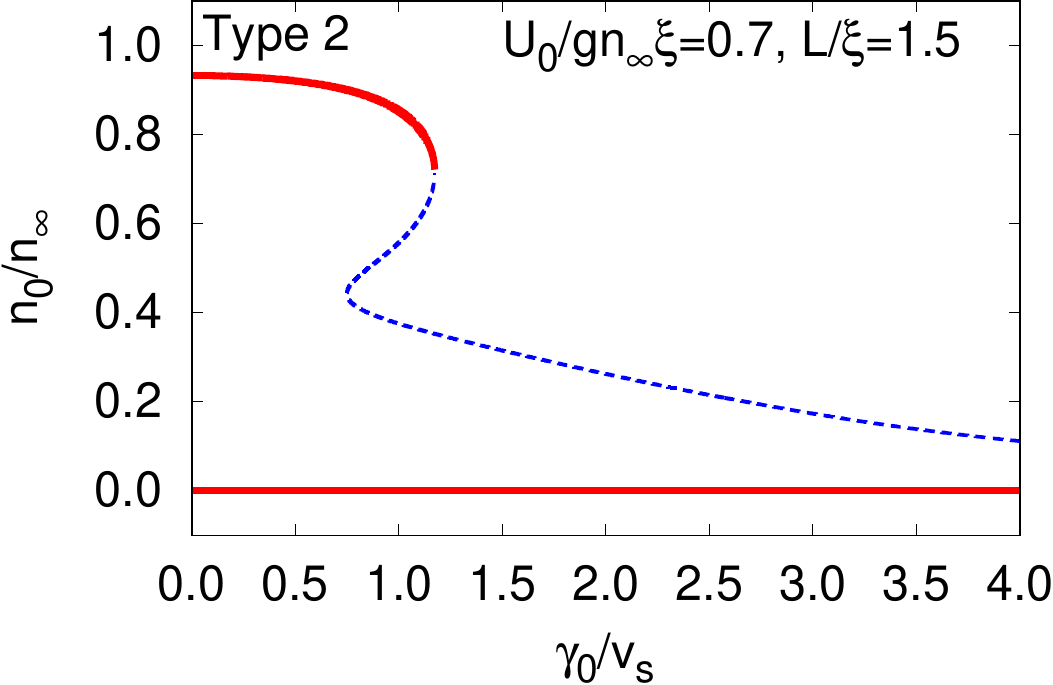}
\caption{$n_0$-$\gamma_0$ diagram for $U_0=0.7g n_{\infty}\xi$ and $L=1.5\xi$. The solid red and doted blue lines represent the stable and unstable states, respectively.}
\label{fig:n0-gamma0_U0.7_L1.5}
\end{figure}

Type 2 emerges in the region adjacent to type 1. A typical $n_0$-$\gamma_0$ diagram is shown in Fig.~\ref{fig:n0-gamma0_U0.7_L1.5}. In type 2, we can see the discontinuous jump between the upper branch and the lower branch. A similar discontinuous jump has been observed in experiments \cite{Labouvie2016}. In contrast, there is no discontinuous jump in the absence of potential barriers (see Fig.~\ref{fig:density_at_origin_U0}). This result means that the discontinuous jump is due to the effects of potential barriers.

\begin{figure}[t]
\centering
\includegraphics[width=8.5cm]{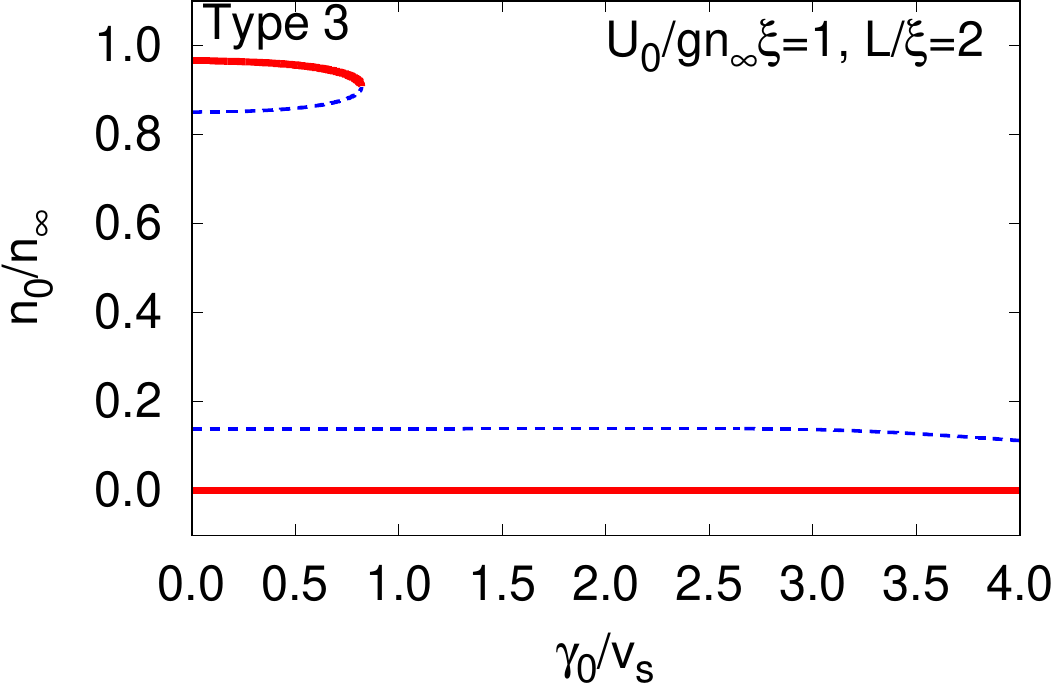}
\caption{$n_0$-$\gamma_0$ diagram for $U_0=1g n_{\infty}\xi$ and $L=2\xi$. The solid red and doted blue lines represent the stable and unstable states, respectively.}
\label{fig:n0-gamma0_U1_L2}
\end{figure}

We show a typical $n_0$-$\gamma_0$ diagram of type 3 in Fig.~\ref{fig:n0-gamma0_U1_L2}. In type 3, the upper and lower branches are completely separated. We can see a saddle-node bifurcation in the upper branch, in which two fixed points collide with each other and are annihilated \cite{Strogatz}. This behavior is similar to that of Josephson junction systems. Theoretically, these systems have been studied using the GP equation or the Ginzburg-Landau equation with a single potential barrier \cite{Baratoff1970,Sols1994,Hakim1997,Kovrizhin2001,Pham2002,Pavloff2002,Kagan2003,Astrakharchik2004,Seaman2005,Bilas2005,Danshita2006,Danshita2007,Watanabe2009,Sykes2009,Takahashi2009,Piazza2010,Kato2010,Watabe2013,Cominotti2014}. In fact, our system can be regarded as a connection of two reverse Josephson junction systems via local loss. The upper branch is reflected by the properties of the Josephson junction, i.e., superfluidity.

\begin{figure}[t]
\centering
\includegraphics[width=8.5cm]{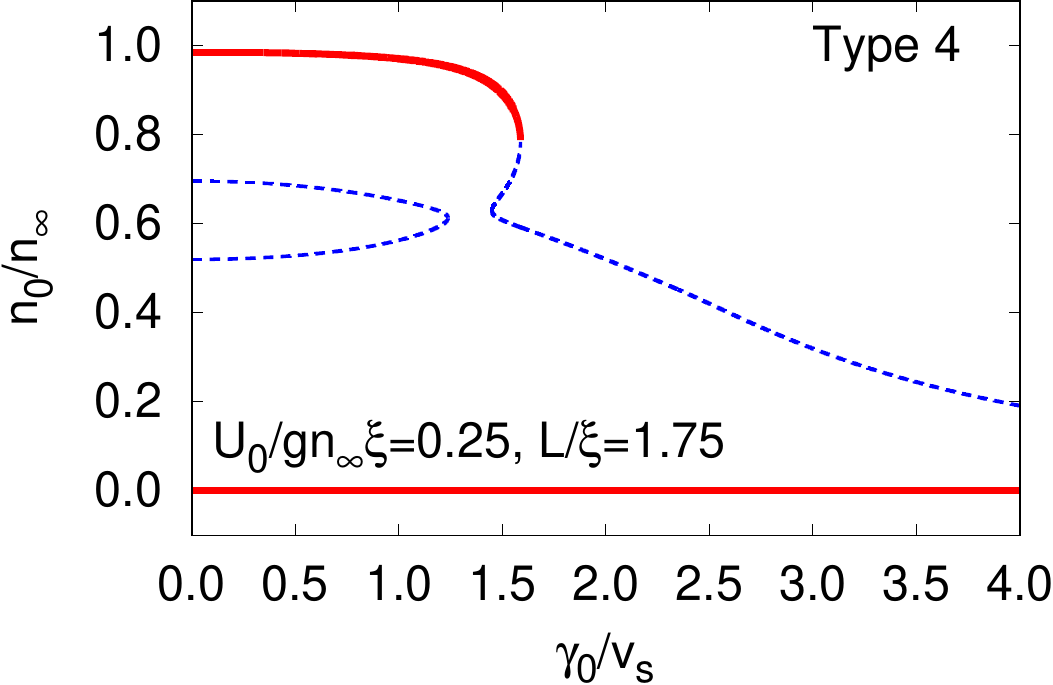}
\caption{$n_0$-$\gamma_0$ diagram for $U_0=0.25g n_{\infty}\xi$ and $L=1.75\xi$. The solid red and doted blue lines represent the stable and unstable states, respectively.}
\label{fig:n0-gamma0_U0.25_L1.75}
\end{figure}

Type 4 emerges in a narrow region surrounded by types 2, 3, and 5 (see inset in Fig.~\ref{fig:state_phase_diagram}). A typical $n_0$-$\gamma_0$ diagram is shown in Fig.~\ref{fig:n0-gamma0_U0.25_L1.75}. In type 4, the upper and lower branches are similar to those of type 2 and one additional branch emerges between the upper and the lower branches.

\begin{figure}[t]
\centering
\includegraphics[width=8.5cm]{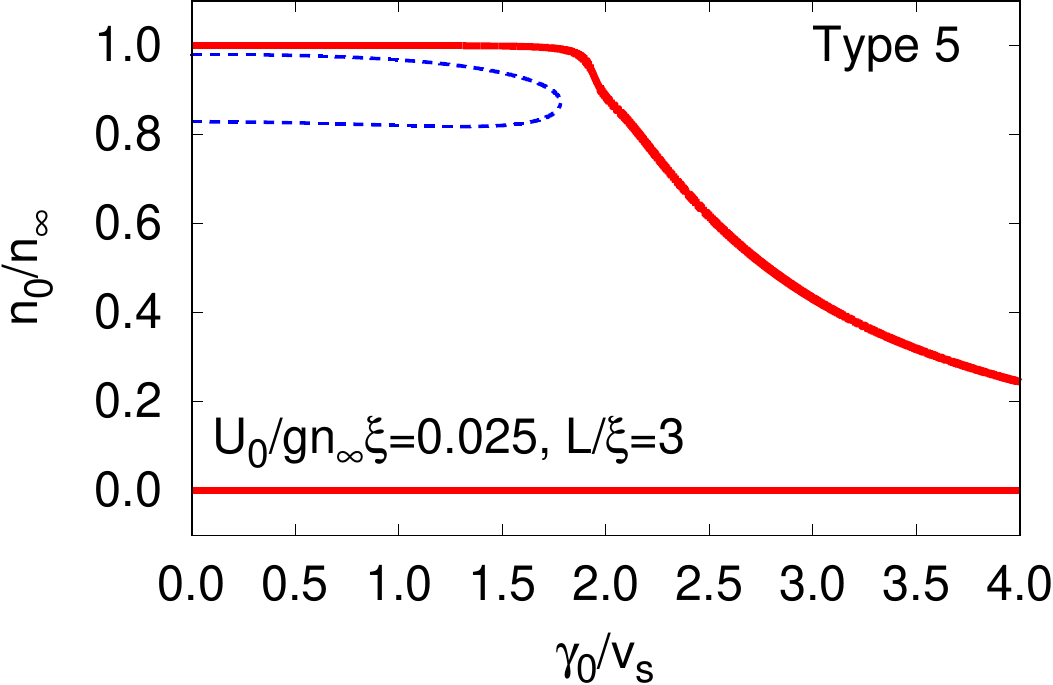}
\caption{$n_0$-$\gamma_0$ diagram for $U_0=0.025g n_{\infty}\xi$ and $L=3\xi$. The solid red and doted blue lines represent the stable and unstable states, respectively.}
\label{fig:n0-gamma0_U0.025_L3}
\end{figure}

A typical $n_0$-$\gamma_0$ diagram of type 5 is shown in Fig.~\ref{fig:n0-gamma0_U0.025_L3}. Type 5 is located between type 1 and type 4. Type 5 is similar to type 4 except for the upper branch. The upper branch of type 5 is similar to that of type 1.

From the above results, we can see the bistability for the whole $\gamma_0$ region in types 1 and 5 and partial regions in types 2, 3, and 4. The difference between the presence and the absence of potential barriers is the existence of the discontinuous jump, which can be seen in types 2, 3, and 4. 

Comparing our results with the experimental ones, we find that our results are in part inconsistent with the experiment \cite{Labouvie2016}. In the small-t$\gamma_0$ region, while only one stable state was observed in the experiment, there are two stable states in our model, one of which is the DS state. One possible reason for this discrepancy is that the local particle loss is modeled as a delta-function form.

\subsection{Anomalous hysteresis}\label{subsec:Anomalous hysteresis}

\begin{figure}[t]
\centering
\includegraphics[width=8.5cm]{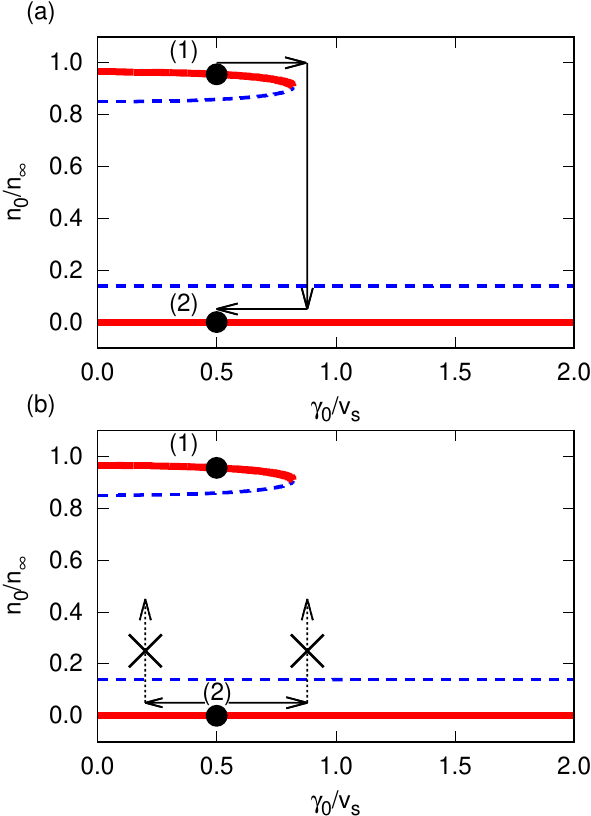}
\caption{Anomalous hysteresis process. The parameters are the same as in Fig.~\ref{fig:n0-gamma0_U1_L2}.}
\label{fig:anomalous_hysteresis}
\end{figure}%

In addition to the bistability, the present system exhibits a nontrivial hysteresis phenomenon, which is called anomalous hysteresis \cite{Yamamoto2012,Yamamoto2013_2,Yamamoto2013}. A feature of anomalous hysteresis is unidirectionality. In conventional hysteresis phenomena, if we observe a discontinuous jump from an initial phase to another phase upon changing the parameters sufficiently slowly, another jump going back to the initial phase exists along the reverse path in the parameter space. However, in anomalous hysteresis, the discontinuous jump exists only in one direction. This phenomenon has been predicted in quantum phase transitions of dipolar or multicomponent Bose gases in an optical lattice \cite{Yamamoto2012,Yamamoto2013} and frustrated magnets \cite{Yamamoto2013_2}, and it can be understood within the framework of the Ginzburg-Landau theory.

Here, we focus on type 3. The processes of anomalous hysteresis are shown in Fig.~\ref{fig:anomalous_hysteresis}. First, we prepare the initial state at point (1) shown in Fig.~\ref{fig:anomalous_hysteresis} (a). Then we increase $\gamma_0$ sufficiently slowly. When the dissipation strength reaches the critical value, the discontinuous jump occurs from the upper branch to the lower branch. After the discontinuous jump, we decrease the dissipation strength and, finally, reach point (2) shown in Fig.~\ref{fig:anomalous_hysteresis} (a). Next, let us consider the inverse process; that is, the initial state is point (2) in Fig.~\ref{fig:anomalous_hysteresis} (b) and the goal is point (1) in Fig.~\ref{fig:anomalous_hysteresis} (b). However, this process is impossible because the lowest branch is stable for the whole $\gamma_0$ region. This means that we cannot reach point (1) starting from point (2) as long as we consider sufficiently slow changes of the parameters. This is nothing but the anomalous hysteresis phenomenon mentioned above. 

Here, we discuss the time scale of changing parameters. For example, let us consider the case shown in Fig.~\ref{fig:anomalous_hysteresis}. The energy difference between the upper stable branch and the lower stable branch is given by the energy of the dark soliton, which is of the order of the chemical potential. In the actual experiments, the chemical potential is typically of the order of $1\;{\rm kHz}$. The inverse of this energy scale gives us the time scale of changing parameters. Therefore, we should change the parameters within a time scale slower than $1 {\rm ms}$. This condition can be easily satisfied in cold-gas experiments. We also remark on the adiabatic condition of the system. The adiabatic condition of the present system, which corresponds to the condition where no excitation is present, is given by the Bogoliubov excitation. We can roughly estimate the times cale to be $100\; {\rm ms}$ for the system size $O(10\xi)$. This means that Bogoliubov excitations are present in the experiment \cite{Labouvie2016} because the experimental timescale is shorter than $100 {\rm ms}$. Nevertheless, the hysteresis loop can be clearly observed. This indicates that the adiabaticity is not a necessary condition but a sufficient condition for observing the hysteresis loop.

At the end of this section, we discuss the feasibility of observing anomalous hysteresis in experiments. Thus far, anomalous hysteresis has not been observed experimentally for the following reasons. In the case of dipolar or multicomponent Bose gases in an optical lattice \cite{Yamamoto2012,Yamamoto2013}, the temperature in the optical lattice has not been lowered enough to observe anomalous hysteresis. In the case of frustrated magnets \cite{Yamamoto2013_2}, it is difficult to tune the parameters to the optimal values for observing anomalous hysteresis. 

In contrast to the previous works, there is no difficulty with our model in achieving sufficiently low temperatures and optimal values of the parameters. However, anomalous hysteresis has not been observed in experiments \cite{Labouvie2016}. There are a few possible reasons for this discrepancy. One is the effects of the harmonic trap. The presence of the trap potential may affect the hysteresis because it changes the boundary condition of the system. In our systems, we fix the wave function as the plane wave at $\infty$. This means that the particles are provided from the bath. This situation is different from the experimental setup, which is isolated from the environment except for the local loss. We also remark that this boundary condition produces an additional nonlinearity. The combination of the boundary conditions at $\infty$ and at the origin determines the velocity at $\infty$ [see Eq.~(\ref{eq:expression_v_infty})]. The velocity depends on the density at the origin. This constraint does not exist in the experimental setup. This difference may affect the existence of anomalous hysteresis. Another one is the effects of optical lattices. The optical lattice extends over the entire system. In contrast to this, in our system, the double delta potentials are localized near the center of the system. This difference may affect the hysteresis. In addition to these points, the width of the local dissipation may affect the stability as discussed in Sec.~\ref{subsec:presence_of_pinning_potential}.

\section{Summary and future prospects}\label{sec:summary}
We have investigated the stability of a BEC with a local one-body loss in double potential barriers by using the mean-field approximation. We obtained the exact solutions of the GP equation in the presence of delta-function potentials with the pure imaginary and real coefficients, which are written by the Jacobi elliptic functions. We showed that there is a wide parameter region, in which two nonequilibrium steady states are dynamically stable, i.e., our model exhibits bistability. We also found the anomalous hysteresis phenomenon in our system.

As a future plan, we will investigate the effects of the width of the local dissipation and the optical lattice potentials. These effects may change the stability of the present system. By studying these effects, we may clarify the origin of the bistability observed in the experiment.

It is interesting to extend our analysis to strongly correlated regimes. Our model is based on the mean-field theory, which can be justified only in weakly correlated regimes. Strongly correlated nonequilibrium states are one of the most difficult problems in various fields. As a topic related to bistability, negative differential conductivity is theoretically studied by using anti-de Sitter space and conformal field theory correspondence \cite{Nakamura2012}.

Another extension is to consider the effects of local multi-body losses, for example, two-body and three-body losses. Particularly, controllable global two-body losses have been realized using the photo-association laser \cite{Tomita2017}. By developing this kind of experimental technique, controllable local two-body losses will be experimentally realized.

\begin{acknowledgments}
M. K. thanks D. A. Takahashi for his lecture on the Jacobi elliptic function, S. Takada for useful discussion, and T. Tomita for useful comments. M. K. was supported by Grant-in-Aid for JSPS Research Fellow No.~JP16J07240. I. D. was supported by JSPS KAKENHI Grants No.~15H05855, No.~25220711, No.~18K03492, and No.~18H05228, by a research grant from CREST, JST, and by the Q-LEAP program of MEXT, Japan.
\end{acknowledgments}


\appendix

\section{STABILITY ANALYSIS}\label{app:stability_analysis}
Here, we explain how to perform the stability analysis of the stationary state. To do this, we investigate real-time dynamics. However, we do not use Eq.~(\ref{eq:time-dependent_GP_eq})  because of some technical reasons described below.

The original problem is defined by an infinite-sized system. However, this system is not tractable numerically. Instead of considering the infinite systems, we consider the finite-size system $(-L_{\rm s}, +L_{\rm s})$, where we take $L_{\rm s}$ to be about $100\xi$. The equation considered here is given by
\begin{align}
i\hbar\frac{\partial}{\partial t}\psi(x, t)&=[1-i\Gamma(x)]\mathcal{L}(x, t)\psi(x, t),\label{eq:real_time_dynamics_GP}\\
\mathcal{L}(x, t)&\equiv -\frac{\hbar^2}{2M}\frac{\partial^2}{\partial x^2}+U(x)-\mu(t)+g|\psi(x, t)|^2,\label{eq:definition_of_L}\\
\mu(t)&\equiv g n_{\infty}+\frac{1}{2}Mv(t)^2,\label{eq:definition_of_time_dependent_mu}\\
v(t)&\equiv \frac{1}{2}\frac{n(0, t)}{n_{\infty}}\gamma_0,\label{eq:definition_of_velocity}\\
\Gamma(x)&\equiv 2+\tanh\left(\frac{x-L_{\rm d}}{W}\right)-\tanh\left(\frac{x+L_{\rm d}}{W}\right),\label{eq:definition_of_Gamma}
\end{align}
where we have introduced the spatially varying dissipation term $\Gamma(x)$. The reason we introduce the dissipation term is to avoid effects of the reflection of the boundary, which does not exist in the original problem. The functional form of the dissipation $\Gamma(x)$ is the same as that used in Ref.~\cite{Reeves2015}. The parameters are set to $L_{\rm d}=L_{\rm s}/2$ and $W=10\xi$. We note that the choice of these parameters is insensitive to the results as long as $L_{\rm d}, W\gg \xi$ are satisfied. We also introduce the time dependence of the chemical potential to converge to the stationary solution at the long time. The boundary condition at the edge of the system is given by
\begin{align}
\left.\frac{\partial\psi(x, t)}{\partial x}\right|_{x=\pm L_{\rm s}}&=\mp i\frac{M v(t)}{\hbar}\psi(\pm L_{\rm s}, t).\label{eq:boundary_condition_numerical_calculations}
\end{align}

We numerically solve Eq.~(\ref{eq:real_time_dynamics_GP}) by using the fourth-order Runge-Kutta method. The centered difference method is used for the space discretization. We use the number of meshes $N_x=2001$-$64001$. In this calculation, we approximate the delta function as the Kronecker $\delta(x-x_j)\simeq (1/\Delta x)\delta_{i,j}$, where $x_i\equiv \Delta x\times i \;[i=-(N_x-1)/2,\cdots, +(N_x-1)/2]$ and $\Delta x$ is the mesh size. We write the discretized wave function at mesh $i$ and time $t$ as $\psi_i(t)$. We have checked that the analytically obtained stationary solutions and the numerically obtained stationary solutions are in good agreement.

The procedure of the stability analysis is as follows. We use the initial conditions as the exact solution plus small random noise. That is, the initial condition is given by $\psi_j(0)=\psi_{\rm exact}(x_j)+\epsilon_j^{\rm R}+i\epsilon_j^{\rm I}$, where $\psi_{\rm exact}(x_j)$ is the exact solution at mesh $j$ and $\epsilon_j^{\rm R}$ and $\epsilon_j^{\rm I}$ are real values. We set $-10^{-4}\le \epsilon_j^{\rm R}, \epsilon_j^{\rm I}\le 10^{-4}$. Then we numerically calculate the real-time dynamics. After long-time evolution [typically $1000\tau\sim 10000\tau$, where $\tau\equiv \hbar/(g n_{\infty})$], we compare the final state with the initial state.

\begin{figure}[t]
\centering
\includegraphics[width=8.5cm,clip]{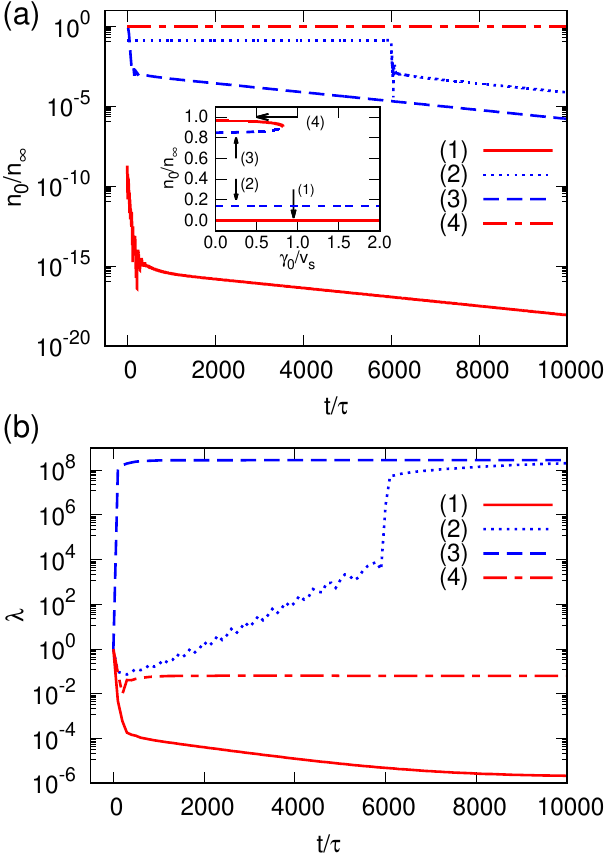}
\caption{(a) Time evolution of the type 3 $n_0$ for $U_0=1g n_{\infty}\xi$, $L=2\xi$, and $\gamma_0=0.2v_{\rm s}$. Red (blue) curves represent stable (unstable) states. Inset: Correspondence of the results, (1)-(4), with the type 3 $n_0$-$\gamma_0$ diagram. (b) Time evolution of $\lambda$. The parameters are the same as in (a).}
\label{fig:time-evolution}
\end{figure}%

A typical example of the time evolution is shown in Fig.~\ref{fig:time-evolution} (a). We see the dynamics of $n_0(t)$ for type 3. We can see that the lowest branch (1) and uppermost branch (4), shown by red lines, are stable against a small perturbation in the initial states. On the other hand, branches (2) and (3), shown by blue lines, are unstable. The instability sets in at $t\sim 6000\tau$ for branch (2) and $t\sim 20\tau$ for branch (3), respectively. In order to quantify the instability, we calculate the quantity \cite{Cassidy2009}:
\begin{align}
\lambda(t)&\equiv \frac{\sum_i|\psi_i(t)-\psi_{\rm exact}(x_i)|^2}{\sum_i|\psi_i(0)-\psi_{\rm exact}(x_i)|^2},\label{eq:difference_norm}
\end{align}
where $\psi_i(t)$ is the wave function at mesh $i$ at time $t$. When $\lambda(t)$ becomes exponentially large, dynamical instability occurs. Figure.~\ref{fig:time-evolution} (b) shows the time evolution of $\lambda(t)$ for the same parameter as in Fig.~\ref{fig:time-evolution} (a). The results show that the values of $\lambda(t)$ for branches (1) and (4) are less than $1$ at all times, while those for branches (2) and (3) are exponentially large after the instability occurs. From these results, we can conclude that branches (1) and (4) are stable and branches (2) and (3) are unstable. In the same manner, we can judge the stability of the exact solutions with other parameters.

\section{DETAILS OF THE DERIVATION OF THE EXACT SOLUTIONS FOR THE EVEN-FUNCTION CASE}\label{app:derivation}
In this Appendix, we describe the details of the derivation of the exact solutions for the even-function case. As we described in Sec.~\ref{subsec:presence_of_pinning_potential}, it is sufficient to consider only the region of $x>0$.

First, we derive Eq.~(\ref{eq:expression_v_infty}). From boundary condition (\ref{eq:boundary_condition_at_infinity}) and the equation of continuity (\ref{eq:equation_of_continuity}), we obtain the current density in stationary states as
\begin{align}
J(x)=-{\rm sgn}(x)n_{\infty}v_{\infty},\label{eq:current_density}
\end{align}
where ${\rm sgn}(\cdot)$ is the sign function. The boundary condition due to the local loss potential (\ref{eq:boundary_condition_loss_term}), can be written as
\begin{align}
\left.\frac{d n(x)}{dx}\right|_{x=+0}&=0,\quad -\frac{\hbar^2}{M}\left.\frac{d\varphi(x)}{d x}\right|_{x=+0}=\frac{\hbar\gamma_0}{2},\label{eq:boundary_condition_density_and_phase_loss_even}
\end{align}
where we have used the assumption of an even function. Using the second Eq.~(\ref{eq:boundary_condition_density_and_phase_loss_even}) and the expression of the current density 
\begin{align}
J(x=+0)=\frac{\hbar}{M}n(0)\left.\frac{d\varphi(x)}{dx}\right|_{x=+0}=-n_{\infty}v_{\infty},\label{eq:current_density_present_case}
\end{align}
we obtain Eq.~(\ref{eq:expression_v_infty}); $v_{\infty}=(n_0/n_{\infty})\gamma_0/2$.

Then we consider solving the GP equation. We define 
\begin{align}
C(x)&\equiv \frac{\hbar^2}{2M}\left|\frac{d\Psi(x)}{d x}\right|^2+\mu|\Psi(x)|^2-\frac{g}{2}|\Psi(x)|^4.\label{eq:definition_of_C_of_x}
\end{align}
It can be easily shown that $C(x)$ is a constant for $0\le x\le L$ and $x>L$. Substituting $\Psi(x)=\sqrt{n(x)}e^{i\varphi(x)}$ and $J(x)$ into Eq.~(\ref{eq:definition_of_C_of_x}), we obtain
\begin{align}
&\frac{\hbar^2}{4M g}\left[\frac{d n(x)}{d x}\right]^2\notag \\
&=n(x)^3-\frac{2\mu}{g}n(x)^2+\frac{2C(x)}{g}n(x)-\frac{M}{g}J(x)^2.\label{eq:rewrite_momentum_flux_tensor}
\end{align}

Here, we consider the outside region ($x>L$). In this region, we obtain $C(x)=(1/2)g n_{\infty}^2+Mv_{\infty}^2n_{\infty}$ from the boundary condition at $x\rightarrow \infty$ (\ref{eq:boundary_condition_at_infinity}). Equation (\ref{eq:rewrite_momentum_flux_tensor}) in the outside region reduces to 
\begin{align}
&\frac{\xi^2}{4}\left[\frac{d n_{\rm out}(x)/n_{\infty}}{d x}\right]^2\notag \\
&=\left[\frac{n_{\rm out}(x)}{n_{\infty}}-1\right]^2\left[\frac{n_{\rm out}(x)}{n_{\infty}}-\left(\frac{v_{\infty}}{v_{\rm s}}\right)^2\right].\label{eq:GP_equation_reduces_to_elliptic_function_form}
\end{align}

\begin{figure}[t]
\centering
\includegraphics[width=8.0cm,clip]{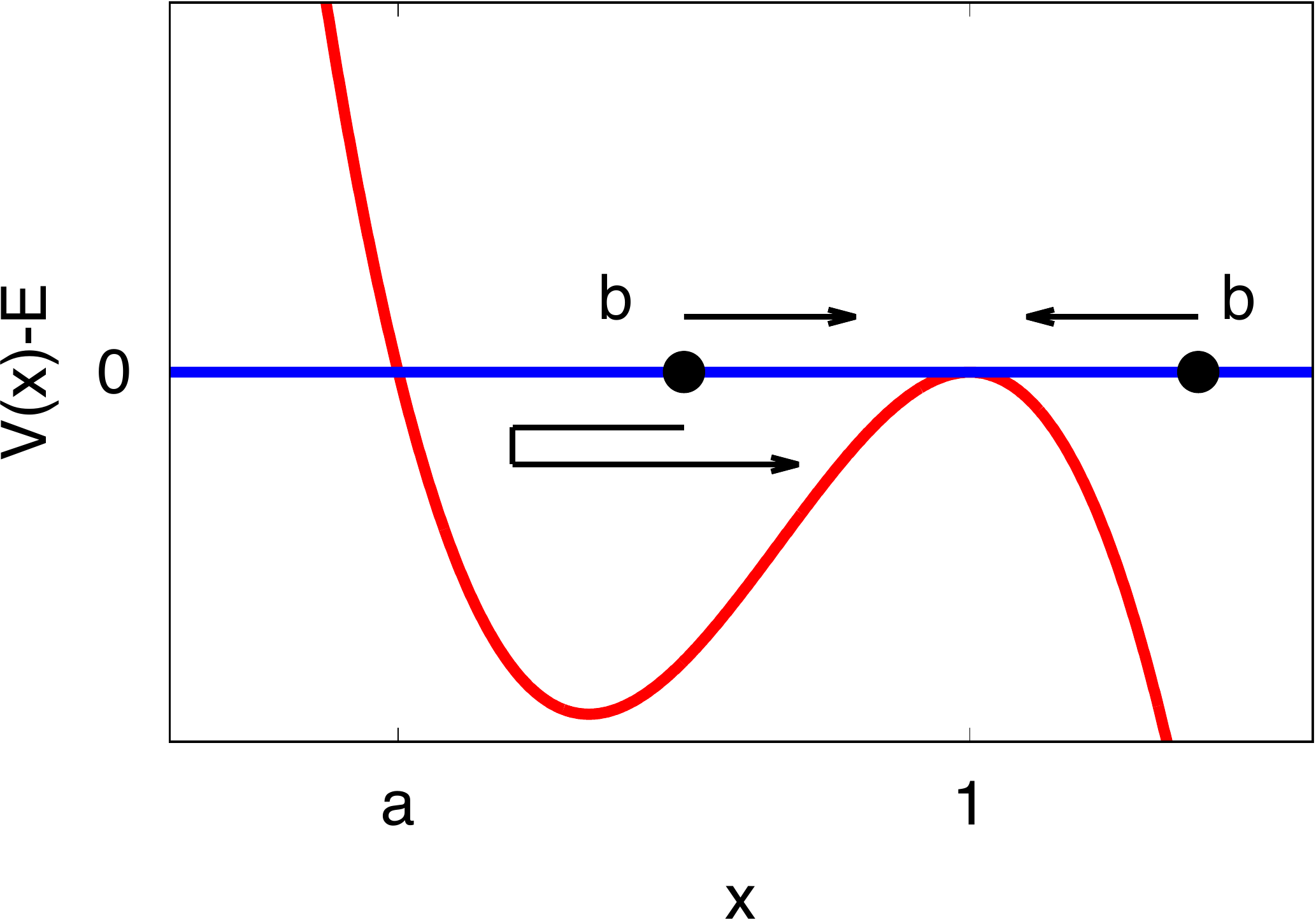}
\caption{Schematic of the motion in the potential $V(x)$. The motion is possible in the region $V(x)-E\le 0$. Arrows indicate the directions of the motion.}
\label{fig:potential_outside}
\end{figure}%

We mention that Eq.~(\ref{eq:GP_equation_reduces_to_elliptic_function_form}) is related to the problem of classical mechanics. Here, we consider a classical particle under the potential $V(x)$. In this case, the energy of the system is given by
\begin{align}
&\frac{1}{4}\left[\frac{d x(t)}{d t}\right]^2+V(x(t))=E\notag \\
\Rightarrow\quad&\frac{1}{4}\left[\frac{d x(t)}{d t}\right]^2=E-V(x(t)),\label{eq:analogy_Newton_equation_of_motion}
\end{align}
where $x(t)$ is the position of the classical particle at time $t$, we set the mass of the particle to $m=1/2$, and $E$ is the total energy. When $V(x)-E=-(x-1)^2(x-a)$\; $(0<a<1)$, this equation has the same structure as Eq.~(\ref{eq:GP_equation_reduces_to_elliptic_function_form}). We assume that $x(t\to\infty)=1$, which corresponds to the boundary condition of the density $n_{\rm out}(x)/n_{\infty}\to 1$ at $x\to\infty$. From this, we can obtain the information on the motion under the potential $V(x)$ in an intuitive way. Figure \ref{fig:potential_outside} shows the potential. From Eq.~(\ref{eq:analogy_Newton_equation_of_motion}), the motion is possible only if $V(x)-E\le 0$. Here, we set the initial condition $x(0)=b$. When $b\ge 1$, we obtain $dx(t)/d t\le 0$. When $b<1$, we have two cases: one is that $x(t)$ is monotonically approaching $1$, and the other is that $x(t)$ is bounced at $x(t)=a$ and goes to $1$. The difference comes from the sign of the initial condition $dx(t)/d t |_{t=0}$.

From the above discussion, we can expect that there are three types of solutions in the outside region. From Eq.~(\ref{eq:GP_equation_reduces_to_elliptic_function_form}) we obtain
\begin{align}
\pm\frac{1}{2}\int^{n(x)/n_{\infty}}_{n_L/n_{\infty}}dX\frac{1}{|1-X|\sqrt{X-(v_{\infty}/v_{\rm s})^2}}=\frac{x-L}{\xi}.\label{eq:integrate_GP_1}
\end{align}
Here, we consider the case $n_L/n_{\infty}<1$. In this case, we can show $n_L\le n_{\rm out}(x)\le n_{\infty}$ from Eq.~(\ref{eq:GP_equation_reduces_to_elliptic_function_form}) and perform the integral in Eq.~(\ref{eq:integrate_GP_1}); then we obtain  Eq.~(\ref{eq:solution_outside}),
\begin{align}
\frac{n_{\rm out}(x)}{n_{\infty}}&=\left(\frac{v_{\infty}}{v_{\rm s}}\right)^2+\left[1-\left(\frac{v_{\infty}}{v_{\rm s}}\right)^2\right]\notag \\
&\times \tanh^2\left[\sqrt{1-\left(\frac{v_{\infty}}{v_{\rm s}}\right)^2}\frac{x-L+ x_{+}}{\xi}\right],\label{eq:solution_outside_app}\\
\frac{x_+}{\xi}&=\frac{1}{\sqrt{1-\left(\dfrac{v_{\infty}}{v_{\rm s}}\right)^2}}\notag \\
&\times \tanh^{-1}\left[\sqrt{\frac{n_L/n_{\infty}-(v_{\infty}/v_{\rm s})^2}{1-(v_{\infty}/v_{\rm s})^2}}\right],\label{eq:definition_of_x_plus_outside_app}
\end{align}
To perform the integral, we used the integral formula
\begin{align}
&\int dx\frac{1}{(p x+q)\sqrt{ax+b}}\notag \\
&=\dfrac{1}{\sqrt{(b p-a q)p}}\log\left|\dfrac{p\sqrt{a x+b}-\sqrt{(bp-a q)p}}{p\sqrt{a x+b}+\sqrt{(bp-a q)p}}\right|,\label{eq:integral_formula_for_outside}
\end{align}
where this formula is valid for $(b p-a q)p>0$.
In the case of $n_L/n_{\infty}>1$, we can obtain a different solution, whose functional form is given by replacing $\tanh$ with $\coth$ in Eq.~(\ref{eq:solution_outside}). However, we cannot find the parameter region where this solution satisfies the boundary conditions. Therefore, we do not consider the case $n_L/n_{\infty}>1$ in the text.

The phase of the outside region can be obtained by integrating Eq.~(\ref{eq:current_density_present_case}). Its expression is given by
\begin{align}
\varphi_{\rm out}(x)&=\varphi_L-\frac{M v_{\infty} (x-L)}{\hbar}\notag \\
&\; -\tan^{-1}\left[\frac{G(x+ x_+)}{v_{\infty}/v_{\rm s}}\right]+\tan^{-1}\left[\frac{G(L+ x_+)}{v_{\infty}/v_{\rm s}}\right],\label{eq:simple_form_of_phase_app}\\
G(x)&=\sqrt{1-\left(\frac{v_{\infty}}{v_{\rm s}}\right)^2}\tanh{\left[\sqrt{1-\left(\frac{v_{\infty}}{v_{\rm s}}\right)^2}\frac{x-L}{\xi}\right]},\label{eq:definition_of_function_G_app}
\end{align}
To perform the integral, we used the mathematical formulas:
\begin{align}
\frac{d}{d x}\tan^{-1}[F(x)]=\frac{\dfrac{dF(x)}{dx}}{1+[F(x)]^2},\quad e^{i\tan^{-1}(x)}=\frac{1+ix}{\sqrt{1+x^2}},\label{eq:formula_for_phase_integral}
\end{align}
where $F(x)$ is a smooth function. 

We can obtain the constraint of the velocity $v_{\infty}$ from the above results. From Eq.~(\ref{eq:GP_equation_reduces_to_elliptic_function_form}), $n_{\rm out}(x)/n_{\infty}\ge (v_{\infty}/v_{\rm s})^2$ must hold. Using $n_{\rm out}(x)/n_{\infty}\le 1$, we obtain the relation
\begin{align}
\left(\frac{v_{\infty}}{v_{\rm s}}\right)^2\le 1\quad \Rightarrow\quad \left(\frac{\gamma_0}{v_{\rm s}}\right)^2\le 4\left(\frac{n_{\infty}}{n_0}\right)^2.\label{eq:constraints_of_the_velocity}
\end{align}
This means that the velocity of the stationary solution is always subsonic. This is consistent with the well-known results for the condition of the existence of a gray soliton in uniform systems.

\begin{figure*}[t]
\centering
\includegraphics[width=13cm,clip]{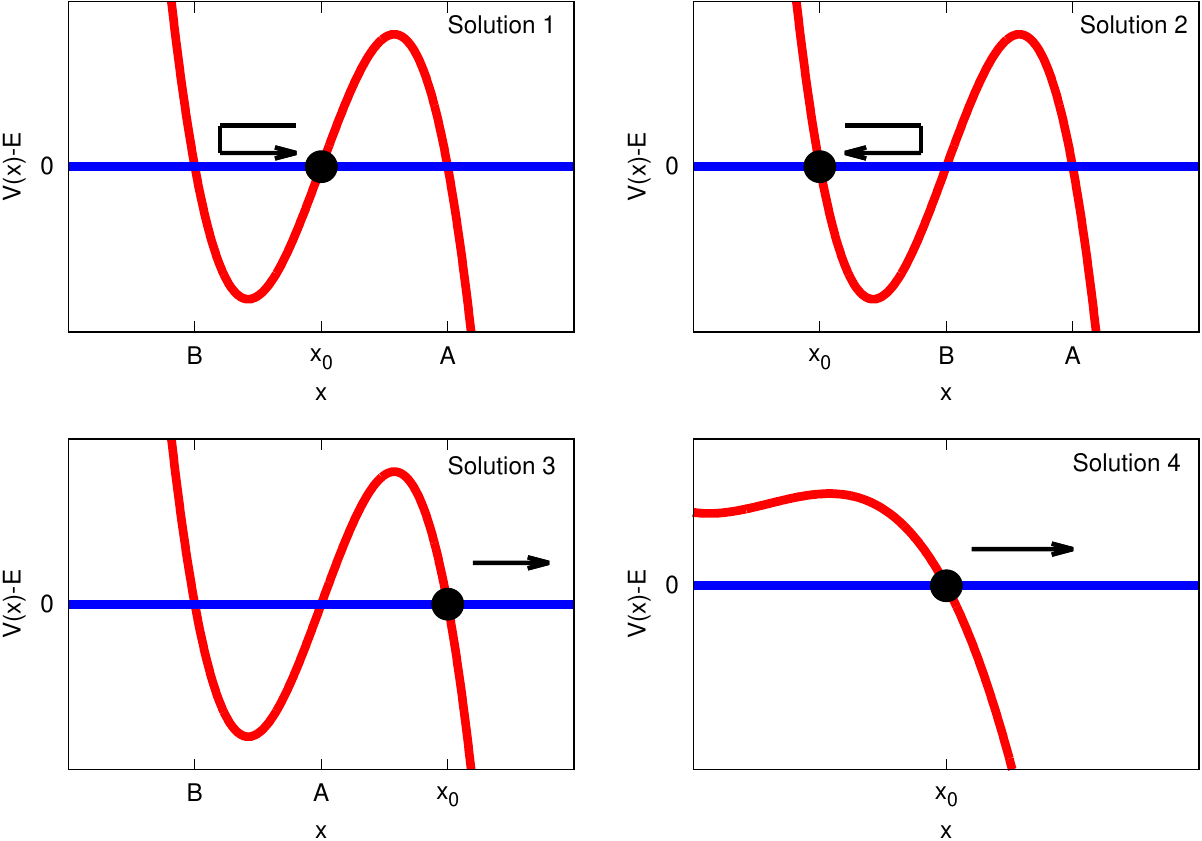}
\caption{Schematic of the motion in the potential $V(x)$. The motion is possible in the region $V(x)-E\le 0$. Arrows indicate the directions of the motion. In solution 4, there is one solution $V(x)-E=0$, hence $A$ and $B=A^{\ast}$ are complex.}
\label{fig:potential_inside}
\end{figure*}

Now, we consider the inside region ($0<x<L$). Using the first Eq.~(\ref{eq:boundary_condition_density_and_phase_loss_even}), Eq.~(\ref{eq:expression_v_infty}), and Eq.~(\ref{eq:rewrite_momentum_flux_tensor}), we can determine $C_{\rm in}\equiv C(x) \;(\text{for }x<L)$ in the inside region:
\begin{align}
\frac{C_{\rm in}}{g n_{\infty}^2}&=\frac{1}{8}\left(\frac{\gamma_0}{v_{\rm s}}\right)^2\frac{n_0}{n_{\infty}}\notag \\
&\quad +\left[1+\frac{1}{8}\left(\frac{\gamma_0}{v_{\rm s}}\right)^2\left(\frac{n_0}{n_{\infty}}\right)^2\right]\frac{n_0}{n_{\infty}}-\frac{1}{2}\left(\frac{n_0}{n_{\infty}}\right)^2.\label{eq:expression_for_Cin_for_gamma_n}
\end{align}
From Eq.~(\ref{eq:expression_for_Cin_for_gamma_n}), we can rewrite (\ref{eq:rewrite_momentum_flux_tensor}) in the inside region as
\begin{align}
&\frac{\xi^2}{4}\left[\frac{d n(x)/n_{\infty}}{dx}\right]^2\notag \\
&=\left[\frac{n(x)}{n_{\infty}}-\frac{n_0}{n_{\infty}}\right]\left[\frac{n(x)}{n_{\infty}}-A\right]\left[\frac{n(x)}{n_{\infty}}-B\right],\label{eq:equation_for_factorized_form}
\end{align}
where $A$ and $B$ were defined by Eqs.~(\ref{eq:expression_of_A_fixed}) and (\ref{eq:expression_of_B_fixed}):
\begin{align}
A&=\frac{1}{2}\left[2+\frac{1}{4}\left(\frac{\gamma_{0}}{v_{\rm s}}\right)^2\left(\frac{n_0}{n_{\infty}}\right)^2-\frac{n_0}{n_{\infty}}+\sqrt{\Delta}\right],\label{eq:expression_of_A_fixed_app}\\
B&= \frac{1}{2}\left[2+\frac{1}{4}\left(\frac{\gamma_{0}}{v_{\rm s}}\right)^2\left(\frac{n_0}{n_{\infty}}\right)^2-\frac{n_0}{n_{\infty}}-\sqrt{\Delta}\right],\label{eq:expression_of_B_fixed_app}\\
\Delta&= \left[\frac{n_0}{n_{\infty}}-2-\frac{1}{4}\left(\frac{\gamma_{0}}{v_{\rm s}}\right)^2\left(\frac{n_0}{n_{\infty}}\right)^2\right]^2-\left(\frac{\gamma_{0}}{v_{\rm s}}\right)^2\frac{n_0}{n_{\infty}}.\label{eq:definition_of_Delta_app}
\end{align}
We can integrate Eq.~(\ref{eq:equation_for_factorized_form}) in a similar manner to the case of the outside region. The corresponding potential of the classical mechanics is given by
\begin{align}
V(x)-E=-(x-x_0)(x-A)(x-B).\label{eq:potential_classical_mechanics_inside_case}
\end{align}
In this case, the initial condition is given by $x(0)=x_0$, which corresponds to $n(x=0)=n_0$. To perform the integral, we need to know the relation between $A$ and $B$. When $\Delta\ge 0$, we obtain $A\ge B$ from Eqs.~(\ref{eq:expression_of_A_fixed}), (\ref{eq:expression_of_B_fixed}), and (\ref{eq:definition_of_Delta}). Therefore, we classify the solutions as four types: solution 1, $\Delta\ge 0,\;B\le x_0\le A$; solution 2, $\Delta\ge 0,\;x_0\le B\le A$; solution 3, $\Delta\ge 0,\;B\le A\le x_0$; and solution 4, $\Delta<0$. The behavior of the potential is shown in Fig.~\ref{fig:potential_inside}.

\begin{figure}[t]
\centering
\includegraphics[width=8.5cm,clip]{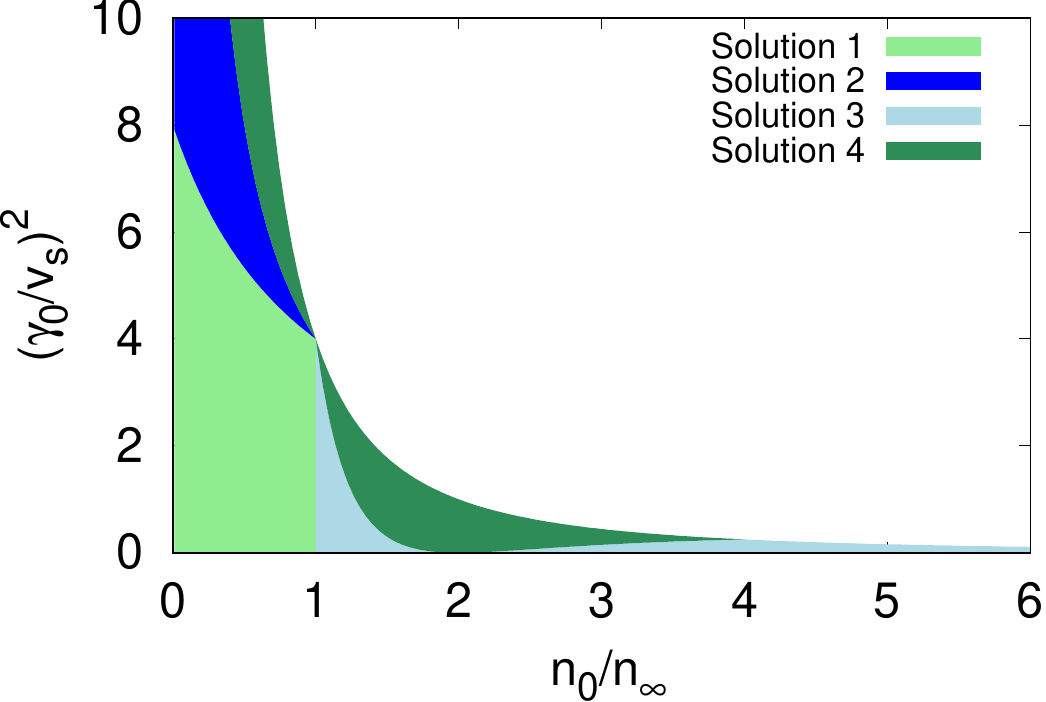}
\caption{Parameter region for each solution.}
\label{fig:parameter_region}
\end{figure}

Here, we consider solution 1. From the inequalities $\Delta\ge 0$, $B\le n_0/n_{\infty}\le A$, and Eq.~(\ref{eq:constraints_of_the_velocity}), this solution exists in the region
\begin{align}
\left(\frac{\gamma_0}{v_{\rm s}}\right)^2\le\frac{8}{1+n_0/n_{\infty}}\text{ and }0\le \frac{n_0}{n_{\infty}}\le 1.\label{eq:condition1_for_solution1}
\end{align}
We plot the parameter region in Fig.~\ref{fig:parameter_region}. Solution1 [Eq.~(\ref{eq:density_solution1})] can be obtained by integration of Eq.~(\ref{eq:equation_for_factorized_form}),
\begin{align}
\frac{n_{\rm in}^{(1)}(x)}{n_{\infty}}&=A-\left(A-\frac{n_0}{n_{\infty}}\right){\rm nd}^2(\Delta^{1/4}x/\xi| m_1),\label{eq:density_solution1_app}\\
\varphi_{\rm in}^{(1)}(x)&=-\frac{1}{2A}\frac{n_0}{n_{\infty}}\frac{\gamma_0}{v_{\rm s}}\frac{x}{\xi}\notag \\
&\;-\frac{1}{2\Delta^{1/4}}\frac{\gamma_0}{v_{\rm s}}\frac{A-n_0/n_{\infty}}{A}\notag \\
&\;\times \Pi[m_1A/(n_0/n_{\infty}); {\rm am}(\Delta^{1/4}x/\xi|m_1) | m_1],\label{eq:phase_for_inside_solution_1_app}\\
m_1&= 1-\frac{A-n_0/n_{\infty}}{\sqrt{\Delta}},\label{eq:m1_solution1_app}
\end{align}
where we have used formula 17.4.63 in Ref.~\cite{Abramowitz_Stegun}. The phase (\ref{eq:phase_for_inside_solution_1_app}), is also obtained by integrating Eq.~(\ref{eq:current_density_present_case}). To perform this, integral, we used the relations:
\begin{align}
\Pi(n; \phi| m)&=\int^{\phi}_0d\theta\frac{1}{(1-n\sin^2\theta)\sqrt{1-m\sin^2\theta}}\notag \\
&=\int^{F(\phi|m)}_0dy\frac{1}{1-n{\rm sn}^2(y|m)},\label{eq:relation_incomplete_integral_of_third_kind}\\
\Pi[n;{\rm am}(x|m)| m]&=\int^x_0dy\frac{1}{1-n{\rm sn}^2(y|m)},\label{eq:relation_incomplite_third_and_amplitude_function}
\end{align}
where $F(\phi|m)$ is the incomplete elliptic integral of the first kind.

The region where solution 2 exists is derived by $\Delta\ge 0,\;n_0/n_{\infty}\le B\le A$, and Eq.~(\ref{eq:constraints_of_the_velocity}):
\begin{align}
&\frac{8}{1+n_0/n_{\infty}}<\left(\frac{\gamma_0}{v_{\rm s}}\right)^2\le \frac{4}{n_0/n_{\infty}}\text{ and }\frac{n_0}{n_{\infty}}\le 1.\label{eq:condition_for_solution2}
\end{align}
The expression of solution 2 is given by
\begin{align}
\frac{n_{\rm in}^{(2)}(x)}{n_{\infty}}&=\frac{n_0}{n_{\infty}}+\left(B-\frac{n_0}{n_{\infty}}\right){\rm sn}^2(\Delta^{1/4}x/\xi | m_2),\label{eq:density_solution2_app}\\
\varphi_{\rm in}^{(2)}(x)&=-\frac{1}{2\Delta^{1/4}}\frac{\gamma_0}{v_{\rm s}}\notag \\
&\times \Pi\left[\left.\frac{B-n_0/n_{\infty}}{n_0/n_{\infty}}; {\rm am}(\Delta^{1/4}x/\xi | m_2)\right|m_2\right],\label{eq:phase_for_solution_2_inside_app}\\
m_2&=\frac{B-n_0/n_{\infty}}{A-n_0/n_{\infty}}.\label{eq:m2_solution2_app}
\end{align}
To obtain Eqs.~(\ref{eq:density_solution2_app}) and (\ref{eq:phase_for_solution_2_inside_app}), we used formula 17.4.62 in Ref.~\cite{Abramowitz_Stegun}.

The region of solution 3 is derived by $\Delta\ge 0,\;B\le A\le n_0/n_{\infty}$, and Eq.~(\ref{eq:constraints_of_the_velocity}):
\begin{align}
&\left(\frac{\gamma_0}{v_{\rm s}}\right)^2\le\frac{4(2-n_0/n_{\infty})^2}{(n_0/n_{\infty})^3}\text{ and }1<\frac{n_0}{n_{\infty}}\le 4,\label{eq:condition1_for_solution3}\\
&\text{or }\left(\frac{\gamma_0}{v_{\rm s}}\right)^2\le\frac{4}{(n_0/n_{\infty})^2}\text{ and }4<\frac{n_0}{n_{\infty}}.\label{eq:condition2_for_solution3}
\end{align}
The expression of solution 3 is given by
\begin{align}
\frac{n^{(3)}_{\rm in}(x)}{n_{\infty}}&=\frac{n_0}{n_{\infty}}+\left(\frac{n_0}{n_{\infty}}-A\right){\rm sc}^2\left(\left.\frac{\Delta^{1/4}}{\sqrt{m_3}}\frac{x}{\xi}\right| m_3\right),\label{eq:solution3_density}\\
\varphi^{(3)}_{\rm in}(x)&=-\frac{\sqrt{m_3}}{2\Delta^{1/4}}\frac{n_0}{n_{\infty}}\frac{\gamma_0}{v_{\rm s}}\left\{\frac{\Delta^{1/4}x}{A\sqrt{m_3}\xi}\right.\notag \\
&\left. +\frac{A-n_0/n_{\infty}}{An_0/n_{\infty}}\right.\notag \\
&\left.\times \Pi\left[\left.A/(n_0/n_{\infty}); {\rm am}(\Delta^{1/4}x/(\sqrt{m_3}\xi) |m_3)\right| m_3\right]\right\},\label{eq:phase_solution_3_inside}\\
m_3&\equiv\frac{\sqrt{\Delta}}{A_1+\sqrt{\Delta}},\quad A_1\equiv -\left(A-\frac{n_0}{n_{\infty}}\right),\label{eq:definitino_of_m3_and_A1}
\end{align}
where ${\rm sc}(x|m)\equiv {\rm sn}(x|m)/{\rm cn}(x|m)$ and we have used formula 17.4.64 in Ref.~\cite{Abramowitz_Stegun}.

The region of solution 4 is derived by $\Delta<0$ and Eq.~(\ref{eq:constraints_of_the_velocity}):
\begin{align}
&\frac{4}{n_0/n_{\infty}}<\left(\frac{\gamma_0}{v_{\rm s}}\right)^2\le\frac{4}{(n_0/n_{\infty})^2}\text{ and }\frac{n_0}{n_{\infty}}\le 1,\text{ or }\label{eq:condition1_for_solution4}\\
&\frac{4(2-n_0/n_{\infty})^2}{(n_0/n_{\infty})^3}<\left(\frac{\gamma_0}{v_{\rm s}}\right)^2\hspace{-0.5em}\le\frac{4}{(n_0/n_{\infty})^2}\text{ and }1<\frac{n_0}{n_{\infty}}\le 4.\label{eq:condition2_for_solution4}
\end{align}
The expression of solution 4 is given by
\begin{align}
\frac{n_{\rm in}^{(4)}(x)}{n_{\infty}}&=\frac{n_0}{n_{\infty}}+A_2\frac{1-{\rm cn}\left(\left.2\sqrt{A_2}\dfrac{x}{\xi}\right|m_4\right)}{1+{\rm cn}\left(\left.2\sqrt{A_2}\dfrac{x}{\xi}\right|m_4\right)}\notag \\
&=\frac{n_0}{n_{\infty}}+A_2{\rm sc}^2(\sqrt{A_2}x/\xi|m_4){\rm dn}^2(\sqrt{A_2}x/\xi|m_4),\label{eq:solution_negative_delta_positive_sign_negative_delta}\\
\varphi_{\rm in}^{(4)}(x)&=-\frac{1}{2\sqrt{A_2}}\frac{n_0}{n_{\infty}}\frac{\gamma_0}{v_{\rm s}}\frac{1}{m_4A_2(C_+-C_-)}\notag \\
&\;\times\left\{(C_+^{-1}-1)\Pi\left[C_+^{-1}; {\rm am}(\sqrt{A_2}x/\xi|m_4)\left.\right|m_4\right]\right.\notag \\
&\left.\quad -(C_-^{-1}-1)\Pi\left[C_-^{-1}; {\rm am}(\sqrt{A_2}x/\xi|m_4)\left.\right|m_4\right]\right\},\label{eq:phase_solution_4}
\end{align}
where we have defined 
\begin{align}
A_2&\equiv \sqrt{2\left(\frac{n_0}{n_{\infty}}\right)^2-\left[2+\left(\frac{v_{\infty}}{v_{\rm s}}\right)^2\right]\frac{n_0}{n_{\infty}}+\left(\frac{v_{\infty}}{v_{\rm s}}\right)^2\frac{n_{\infty}}{n_0}},\label{eq:definitino_of_A2}\\
m_4&\equiv \frac{1}{2A_2}\left[A_2-\frac{3n_0}{2n_{\infty}}+1+\frac{1}{8}\left(\frac{\gamma_0}{v_{\rm s}}\right)^2\left(\frac{n_0}{n_{\infty}}\right)^2\right],\label{eq:expression_of_m_in_terms_of_A2_A1_positive_sign_negative_delta}\\
C_{\pm}&\equiv \frac{1}{2}\left(D\pm\sqrt{D^2+\frac{4n_0/n_{\infty}}{m_4A_2}}\right),\label{eq:definition_of_C_pm_for_solution4}\\
D&\equiv \frac{A_2-n_0/n_{\infty}}{m_4A_2}.\label{eq:definition_of_D_app}
\end{align}
Here, we have used formulas 16.18.4 and 17.4.71 in Ref.~\cite{Abramowitz_Stegun}.

What remains to do is to determine the parameters $n_L$, $\varphi_L$, and $n_0$ by connecting the inside and the outside solutions via the boundary conditions. $n_L$ and $\varphi_L$ are determined by the first expression of Eq.~(\ref{eq:boundary_condition_for_continuity}), that is, $n_L^{(i)}=n_{\rm in}^{(i)}(x=L)$ and $\varphi^{(i)}_L=\varphi_{\rm in}^{(i)}(x=L)$. The explicit expressions for the density are given by
\begin{align}
\frac{n_L^{(1)}}{n_{\infty}}&=A-\left(A-\frac{n_0}{n_{\infty}}\right){\rm nd}^2(\Delta^{1/4}L/\xi| m_1),\label{eq:boundary_solution1}\\
\frac{n_L^{(2)}}{n_{\infty}}&=\frac{n_0}{n_{\infty}}+\left(B-\frac{n_0}{n_{\infty}}\right){\rm sn}^2(\Delta^{1/4}L/\xi | m_2),\label{eq:boundary_solution2}\\
\frac{n_L^{(3)}}{n_{\infty}}&=\frac{n_0}{n_{\infty}}+\left(\frac{n_0}{n_{\infty}}-A\right){\rm sc}^2\left(\left.\frac{\Delta^{1/4}}{\sqrt{m_3}}\frac{L}{\xi}\right|m_3\right),\label{eq:boundary_solution3}\\
\frac{n_L^{(4)}}{n_{\infty}}&=\frac{n_0}{n_{\infty}}+A_2{\rm sc}^2(\sqrt{A_2}L/\xi | m_4){\rm dn}^2(\sqrt{A_2}L/\xi|m_4).\label{eq:boundary_solution4}
\end{align}
$n_0$ is determined by boundary condition (\ref{eq:boundary_condition_loss_term}), which reduces to
\begin{align}
&\dfrac{\hbar^2}{4M}\left[\left.\dfrac{d n(x)}{d x}\right|_{x=L+0}-\left.\dfrac{d n(x)}{d x}\right|_{x=L-0}\right]=U_0n(L),\label{eq:continuity_condition_pinned_potential_density}\\
&\left.\dfrac{d\varphi(x)}{d x}\right|_{x=L+0}=\left.\dfrac{d\varphi(x)}{d x}\right|_{x=L-0}
.\label{eq:continuity_condition_pinned_potential_phase}
\end{align}
Equation (\ref{eq:continuity_condition_pinned_potential_phase}) is automatically satisfied due to the expression of the current density (\ref{eq:current_density_present_case}). Equation (\ref{eq:continuity_condition_pinned_potential_density}) reduces to
\begin{align}
&S_{\rm out}(i)\left|\frac{n_L^{(i)}}{n_{\infty}}-1\right|\sqrt{\frac{n_L^{(i)}}{n_{\infty}}-\frac{1}{4}\left(\frac{\gamma_0}{v_{\rm s}}\right)^2\left(\frac{n_0}{n_{\infty}}\right)^2}\notag \\
&\hspace{1.0em}-S_{\rm in}(i)\sqrt{\left[\frac{n_L^{(i)}}{n_{\infty}}-\frac{n_0}{n_{\infty}}\right]\left[\frac{n_L^{(i)}}{n_{\infty}}-A\right]\left[\frac{n_L^{(i)}}{n_{\infty}}-B\right]}\notag \\
&\hspace{2.0em}=\frac{2 M\xi U_0}{\hbar^2}\frac{n_L^{(i)}}{n_{\infty}},\label{eq:boundary_condition_derivative_of_density_at_L_full_form}\\
&S_{\rm out}(i)\equiv {\rm sgn}\left[\left.\frac{d n_{\rm out}(x)}{d x}\right|_{x=L}\right],\label{eq:definition_of_S_out}\\
&S_{\rm in}(i)\equiv {\rm sgn}\left[\left.\frac{d n_{\rm in}^{(i)}(x)}{d x}\right|_{x=L}\right].\label{eq:definition_of_S_in}
\end{align}
Because $n_L^{(i)}$ is a function of $n_0$, Eq.~(\ref{eq:boundary_condition_derivative_of_density_at_L_full_form}) is a one-variable equation of $n_0$ for fixed $\gamma_0$ and $U_0$. Therefore, the problem of solving the GP equation (nonlinear differential equation) reduces to solving the one-variable equation (\ref{eq:boundary_condition_derivative_of_density_at_L_full_form}). Because we cannot obtain the analytical solutions of Eq.~(\ref{eq:boundary_condition_derivative_of_density_at_L_full_form}), we solve this equation numerically.

As mentioned in the text, we cannot find the parameter region where solutions 3 and 4 satisfy the boundary conditions. This means that $n_0/n_{\infty}$ moves only $0\le n_0/n_{\infty}\le 1$.

\section{DETAILS OF THE DERIVATION OF THE EXACT SOLUTIONS FOR THE ODD-FUNCTION CASE}\label{app:derivation_odd}
In this Appendix, we show the expression of boundary condition (\ref{eq:boundary_condition_pinning_potential}) for the odd-function case. Substituting Eqs.~(\ref{eq:tanh_solution_for_odd_function}), (\ref{eq:inside_solution_for_odd_function}), and (\ref{eq:matching_condition_equation_odd_function}) into Eq.~(\ref{eq:boundary_condition_pinning_potential}), we obtain
\begin{align}
&\frac{e^{i\varphi_0}}{1+m_0}\left[1+m_0-2m_0{\rm sn}^2\left(\left.\sqrt{\frac{2}{1+m_0}}\frac{L}{\xi}\right|m_0\right)\right]\notag \\
&-\frac{2\sqrt{m_0}}{1+m_0}{\rm cn}\left(\left.\sqrt{\frac{2}{1+m_0}}\frac{L}{\xi}\right|m_0\right){\rm dn}\left(\left.\sqrt{\frac{2}{1+m_0}}\frac{L}{\xi}\right|m_0\right)\notag \\
&\quad =\frac{2MU_0\xi}{\hbar^2}\sqrt{\frac{2m_0}{1+m_0}}{\rm sn}\left(\left.\sqrt{\frac{2}{1+m_0}}\frac{L}{\xi}\right|m_0\right),\label{eq:boundary_condition_for_odd_function_case}
\end{align}
where $\varphi_0$ has been determined by Eq.~(\ref{eq:matching_condition_equation_odd_function}).


\end{document}